\documentclass[aps,physrev,reprint,groupedaddress,nofootinbib]{revtex4-2}

\usepackage{graphicx}
\usepackage[hidelinks]{hyperref}
\usepackage{xcolor}
\usepackage{amsfonts}
\usepackage{amssymb}
\usepackage{amsmath}
\usepackage{color}
\usepackage[normalem]{ulem}
\usepackage{comment}

\newcommand{\be}{\begin{equation}}
\newcommand{\ee}{\end{equation}}
\definecolor{darkgreen}{RGB}{27,123,27}

\hypersetup{colorlinks=true,
	linkcolor=blue,
	urlcolor=blue,
	citecolor=blue,
	pdfhighlight=/N
}

\begin{document}

\title{Slow activity decay in excitable models with discontinuous phase transitions}

\author{Paulo H. Lorenzoni$^{1,2}$, G\'eza \'Odor$^2$, Silvio C. Ferreira$^{1,4}$, and R\'obert Juh\'asz$^3$}
\email[]{juhasz.robert@wigner.hun-ren.hu}
\affiliation{$^1$Departamento de Física, Univ. Fed. Viçosa, Viçosa, Minas Gerais, 36570-900, Brazil\\
$^2$HUN-REN Centre for Energy Research, 1121 Budapest, Konkoly Thege str. 29-33.\\
$^3$HUN-REN Wigner Research Centre for Physics, H-1525 Budapest, P.O. Box 49, Hungary\\
$^4$Instituto de Ci\^{e}ncias Matem\'{a}ticas e de Computa\c{c}\~{a}o, Universidade de S\~{a}o Paulo, S\~{a}o Carlos, SP 13566-590, Brazil}

\date{\today}

\begin{abstract}
As a simple model of interacting systems of excitable degrees of freedom, we consider a threshold contact process on various lattices and networks, in which a successful activation event requires the presence of more than one active neighbors. We show by combining  numerical simulations and a phenomenological theory that, in the two-phase coexistence region and for a sufficiently low initial activity, a slower-than-exponential temporal decay of the global density emerges, caused by the spontaneous formation of slowly vanishing and non-communicating clusters of activity. This slow-decay phenomenon is found to appear generally for regular lattices and also for finite-dimensional random lattices such as the Voronoi-Delaunay network. However, the slow-decay is impeded by small-world property, as demonstrated by simulations on random-regular networks. In the case of a power-law decay of the order parameter, which is valid among others on two-dimensional regular lattices, our phenomenological theory points out that the decaying state is stable against unbounded nucleation only if the decay exponent exceeds $1/2$. Besides the well-known Griffiths effects of quenched random systems, this slow-decay phenomenon rooted in the threshold condition provides an alternative mechanism of off-critical but scale-free dynamics, occurring also in the absence of disorder.  
\end{abstract}


\maketitle

\section{Introduction}

Threshold models, introduced to describe excitable dynamics, provide a simple description of complex phenomena 
ranging from brain science~\cite{Greenberg-Hastings,CCdyncikk}, through surface catalytic reactions~\cite{marro2005}, 
cooperative sequential adsorption~\cite{evans_review}, ecological systems with Allee effect ~\cite{sato,Gast-Bea-Alle}, 
fracture phenomena to socio-physics~\cite{Schelling01071971}.
In their simplest form, they are defined on regular lattices, the sites of which are either active or inactive ~\cite{Sakoda01011971}. 
They can also be considered as generalizations of the contact process~\cite{ContactProcess} (CP), in which a single active site can ``infect'' its neighbors, competing with a spontaneous self-deactivation. As a consequence of this competition, various non-equilibrium phase transitions can occur~\cite{marro2005,HHL}. In the high-dimensional mean-field limit, 
discontinuous phase transition happens~\cite{PhysRevE.67.056114}, but  fluctuations can turn this transition into a continuous one in
in lower dimensions~\cite{odorbook}.

Besides fluctuations, another circumstance that is able to change an originally discontinuous transition to a continuous one is quenched heterogeneity~\cite{round}. 
Beyond that, heterogeneity affects not only the critical point but it   
also leads to the formation of an extended phase, the so-called 
Griffiths phase (GP) around the critical point \cite{Griffiths1969}. Here, the system is semi-critical, having a finite length scale but infinite time scale, which manifests itself in anomalous (typically power-law) dynamics.

Here, we show that a similar phenomenon can emerge even in homogeneous systems with discontinuous transitions as a consequence of surface effects of ordered domains at the phase separation. 
We investigate the relaxation dynamics of different variants of the contact process extended with a threshold rule, for which activation of an empty site is allowed only if the number of its active neighbors is greater than a certain threshold $k\ge 1$.  

Equivalent to the threshold models are the quadratic (QCP)~\cite{evans_pre} and higher-order contact processes, with 'A' particle reactions $nA\to (n+l)A$, $mA\to\emptyset$ ($n>m$). 
The $n=2$, $m=1$ QCP is a lattice realization of Schlögl's second model \cite{Schlogl2}. 
In the QCP, spontaneous deactivation competes with a cooperative activation (also called particle creation), which requires the presence of pairs of active neighboring sites (particles).  Unlike the standard contact process, the mean-field limit of these models exhibits a mixed type of phase transition, with a discontinuous jump of the steady-state 
particle density and dynamical scaling of the order parameter~\cite{PhysRevResearch.3.013106}, 
which is also the case on two-dimensional lattices \cite{evans_prl}. 
Note that in $1+1$ dimensions, due to the fluctuations, the QCP has a continuous phase 
transition, belonging to the directed percolation (DP) class~\cite{DKDP} of the CP.

An interesting property of QCP, also present at the mean-field level, is the emergence of an extended bistability or two-phase coexistence domain, in which both the zero-density absorbing steady state and the positive-density steady state are stable against local perturbations realized by inserting droplets of the other phase into the system \cite{evans_prl}. Using the activation rate $\lambda$ as a control parameter and starting the system's evolution from a randomly occupied state with density $\rho_0$, the steady state into which the system settles depends on $\lambda$ and $\rho_0$. If $\lambda<\lambda_c$, where $\lambda_c$ marks the phase transition for $\rho_0=1$, the steady state will be the absorbing one, for any $\rho_0$. 
If, however, $\lambda>\lambda_c$, the system enters the bistability domain in which the steady state depends also on the initial density $\rho_0$. If $\rho_0<\rho_0^c(\lambda)$, where 
$\rho_0^c(\lambda)$ is a $\lambda$-dependent critical value, the steady state will be the absorbing one; otherwise, the steady state is an active state with a positive density $\rho_{\infty}(\lambda)$. 
According to the rigorous results in Ref. \cite{Chen} for square lattices, 
$\rho_0^c(\lambda)>0$ for any $\lambda>\lambda_c$, thus the bistability domain extends to arbitrarily large $\lambda$.

Despite the ubiquitous appearance, the non-mean-field relaxation dynamics of these models on low-dimensional lattices is not so well known. 
The relaxation of QCP on two-dimensional lattices was studied by simulations in Ref. \cite{evans_pre}, however, with no conclusion on the functional form of time-dependence of density $\rho(t)$ in the bistable domain, while, in Ref.~\cite{Chen}, an algebraically decaying upper bound for $\rho(t)$ was proved for the same model for any large $\lambda$ and sufficiently small $\rho_0$. 

In this work, we are interested in the relaxation dynamics of threshold contact processes on various, mainly two-dimensional, regular or random lattices, including also a variant that is only slightly different from the two-dimensional QCP studied in Refs. \cite{evans_prl,evans_pre,Chen}. A common feature of these models is that they all possess a discontinuous phase transition and an extended bistability domain. 
We will show by Monte Carlo simulations and by means of a phenomenological theory that, in the latter domain, the relaxation is generally slower than exponential; for the QCP on the square lattice it is algebraic with non-universal decay exponents.  
This phenomenon is caused ultimately by the fact that the threshold processes under study are reducible Markov processes, where the state space of the infinite system contains an infinite number of subspaces that cannot be left by the dynamics \cite{vankampen}. 
This manifests itself as a low-density initial state that will evolve to a non-communicating set of activity clusters, locally belonging to the active phase. As such they get extinct very slowly, leading to a slower-than-exponential decay of the global density. This phenomenon is reminiscent of the Griffiths phase of the standard contact process with quenched disorder, in which the spatial fluctuation of randomness leads to the emergence of effectively non-communicating, locally supercritical rare regions that are embedded in a subcritical background. In the threshold process, the formation of non-communicating regions of long-lasting activity is caused by the reducibility of the process; therefore, this mechanism does not necessitate the presence of quenched disorder.  
We will go beyond regular lattices to investigate the effects of underlying graphs by relaxing the properties of regularity and locality. In particular, we consider the relaxation
dynamics on Voronoi-Delaunay networks and 
random regular networks. In this way, we show an alternative mechanism of slow, scale-free relaxation dynamics in an extended parameter region, that is different from the Griffiths phase of quenched disorder systems and observable in a broad class of systems.

\section{Models and Methods}
\label{sec:MM}

\subsection{Definition of the model}

\begin{figure}[ht]
    \centering
    \includegraphics[width=0.8\linewidth]{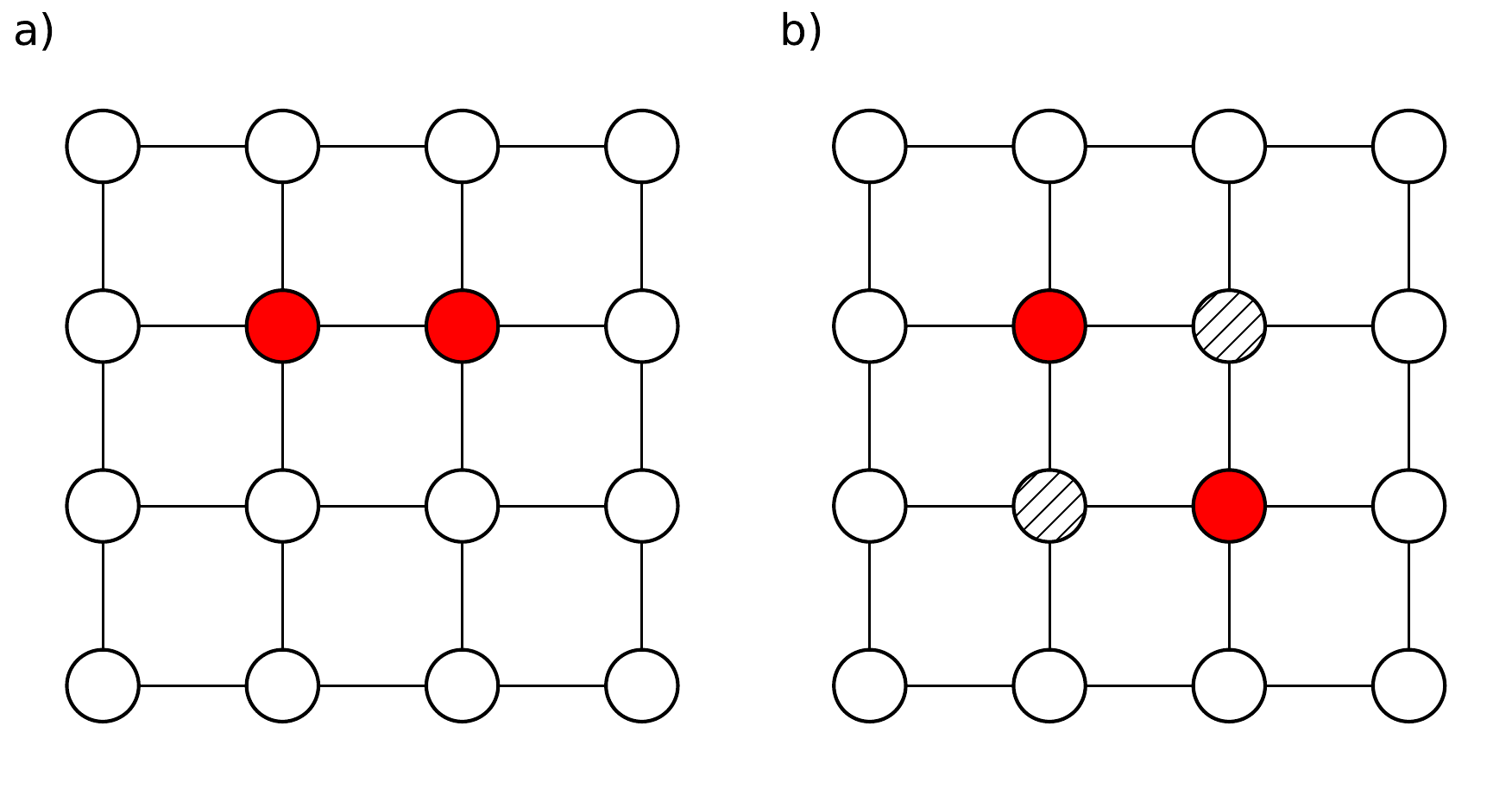}
    \caption{On a square lattice, for a threshold $\Theta=1/2$, an inactive site (white) requires at least two active neighbors (red) to become active. In a) there is no activation possibility. In b) there are two possible activation events in hatched sites.         
    }
    \label{configuration}
\end{figure}

We consider stochastic processes on various types of lattices and networks which are specified by their adjacency matrices, $A_{ij}$. The latter is defined as $A_{ij} =1$ if site $i$ and site $j$ are neighboring and $A_{ij}=0$ otherwise. 
Each site of the lattice can be in one of two states, active or inactive, which may be encoded by a binary variable $\sigma_i$ at site $i$, having values $\sigma_i=1$ and $\sigma_i=0$, respectively. 
On this state space, we consider a continuous-time Markov process with the following independent transitions. First, each active site ($\sigma_i=1$) becomes spontaneously inactive ($\sigma_i=0$) with a rate $\mu$. Without restricting the generality, this rate will be set to unity, $\mu=1$. Second, each active site attempts to activate one of its neighbors with a rate $\lambda$. This is selected randomly with equal probabilities from all of its neighbors. The activation is successful if the chosen site ($j$) is inactive and the following condition is fulfilled:
\be
\frac{\sum_i \sigma_i A_{ij}}{\sum_i A_{ij}} \geq \Theta.
\label{condition}
\ee
In words, the fraction of the number of active neighbors to the degree  (number of all neighbors) of the target site has to be greater than or equal to a predefined threshold $\Theta$. Note that, without this condition, the model reduces to the standard contact process.
Fig. \ref{configuration} illustrates consequences of the threshold condition for the square lattice with $\Theta=1/2$:
Figure~\ref{configuration}a shows a configuration with no possible activation, while Figure~\ref{configuration}b displays two possible activation events. Note that, when the four marked sites are simultaneously active in (b), the activity cannot spread beyond the $2 \times 2$ domain unless there are nearby active sites such that the threshold condition is satisfied. This is one among other details that distinguish the threshold model from the pair-contact process (PCP)~\cite{odorbook}.

\subsection{Mean-field description}

It is instructive to examine the above model within a mean-field approximation that neglects correlation of the occupation variable $\sigma_i$ at different sites. As it turns out, the appearance of a bistability domain is reflected already by this simple description. To illustrate the general phenomenology of this model class, we will discuss here the case of a square lattice with $\Theta=1/2$, i.e., for a successful activation of a vacant site at least two active neighbors are needed. 

\begin{figure}[ht]
    \centering
    \includegraphics[width=1\linewidth]{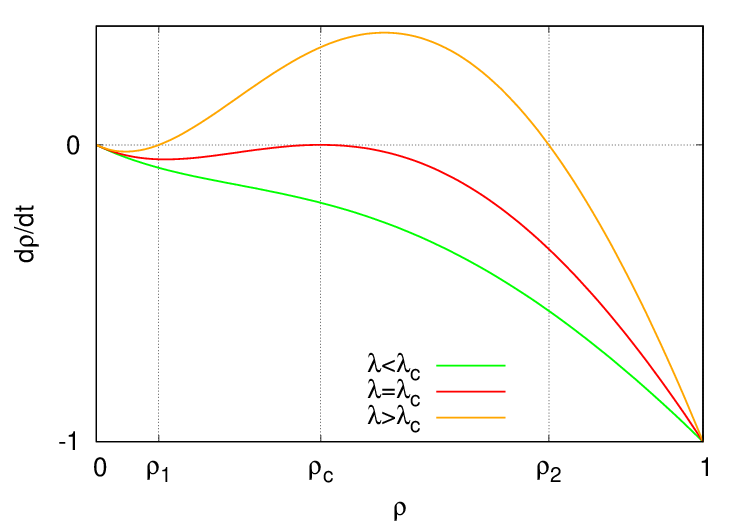}
    \includegraphics[width=1\linewidth]{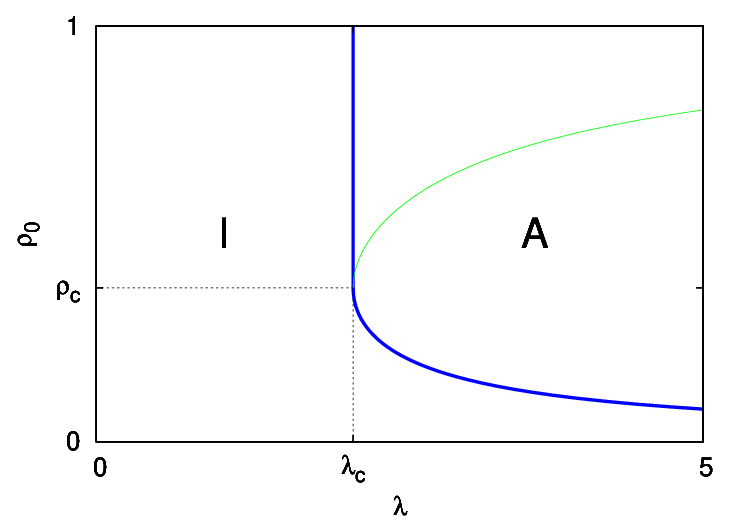}
    \caption{
    \label{fig_mf} Top. Time derivative of the local density $\rho(t)$ as a function of the density, obtained by the mean-field approximation in Eq. (\ref{mf}) for a square lattice with $\Theta=1/2$. The three curves correspond to three different values of $\lambda$, $\lambda=1$, $\lambda=\lambda_c=2^{8/3}/3$, and $\lambda=4$. Bottom. Mean-field phase diagram of the model. The phase boundary between the active (A) and inactive (I) phase is indicated by the thick blue line. The vertical coordinate of the thin green line gives the stationary density within the active phase.}
\end{figure}

Restricting ourselves to homogeneous initial states, the evolution of the local occupation probabilities at time $t$,\
$\rho(t)=\langle\sigma_i\rangle$, where $\langle\cdot\rangle$ denotes an average over stochastic histories, is governed by the equation 
\be
\dot{\rho}(t)=\lambda\rho(1-\rho)[1-(1-\rho)^3]-\rho,
\label{mf}
\ee
where $1-(1-\rho)^3$  is the probability that at least one of the three remaining neighbors, besides the one responsible for the activation, is also active.
The time derivative of $\rho(t)$ is plotted against $\rho$ for different values of $\lambda_c$ in Fig. \ref{fig_mf}. 
In terms of $\lambda$, two domains can be distinguished, which are separated by the critical value $\lambda_c=2^{8/3}/3\approx 2.1165$. 
If $\lambda<\lambda_c$, the fixed point at $\rho=0$ is attractive for any initial density $\rho_0$. At $\lambda=\lambda_c$, a new fixed point appears at $\rho=\rho_c=1-2^{-2/3}\approx 0.3700$, which is attractive for initial densities $\rho_0\ge \rho_c$ but, for $\rho_0<\rho_c$, the zero-density fixed point is still attractive. 
In the domain $\lambda>\lambda_c$, the right-hand-side of Eq. (\ref{mf}) will have two real zeroes, $\rho_1$ and $\rho_2(>\rho_1)$, besides the trivial one at $\rho=0$. As can be read off from Fig. \ref{fig_mf}, the fixed point at $\rho_1$ is unstable while the other one is stable. This means that initial densities with $\rho_0<\rho_1$ are attracted by the zero-density fixed point, whereas for $\rho_0>\rho_1$ the density tends to the stationary value $\rho_2$. 
Thus, we can see that $\lambda_c$ marks the onset of the bistability domain in which the stationary density at late times (either zero or $\rho_2$ depends on the initial condition. 
In this domain, the separatrix $\rho_0^c(\lambda)$ between the zero-density (inactive) and finite-density (active) stationary phase is determined by $\rho_0^c(\lambda)=\rho_1(\lambda)$. 
The phase diagram of the model within the mean-field approximation is shown in Fig. \ref{fig_mf}.
The asymptotic shape of this separatrix at large $\lambda$ is given as 
$\rho_0^c(\lambda)=\frac{1}{3}\lambda^{-1}+O(\lambda^{-2})$. 
The stationary density $\rho_{\infty}(\lambda)=\rho_2(\lambda)$ displays a discontinuous transition at $\lambda_c$, and the density approaching its critical value $\rho_c$ as  $\rho_{\infty}(\lambda)-\rho_c\sim(\lambda-\lambda_c)^{1/2}$. 
Furthermore, at $\lambda=\lambda_c$, if the initial density is above $\rho_c$, the density approaches its stationary value according to a power law, $\rho(t)-\rho_c\sim t^{-1}$; otherwise, the approach to the stationary density is exponential. 

\subsection{Numerical simulations of the threshold model}

We performed simulations of the above-defined threshold model in the following way. 
First, an active site is chosen randomly with uniform probability.
Then, with a probability $\mu/(\mu+\lambda)$, this site becomes spontaneously inactive or, with the complementary probability $\lambda/(\mu+\lambda)$, one of its nearest neighbors ($j$) is picked with equal probabilities. If the selected neighbor $j$ is inactive and the threshold condition is satisfied (\ref{condition}), it is made active. 
The time increment assigned to this update is  $\Delta t = -\ln z / (\mu N+\lambda N)$, where $z$ is a random number 
with a uniform distribution between (0, 1) and $N$ is the number of active sites. The dynamics is implemented using the optimized Gillespie algorithm (OGA)~\cite{COTA2017303}.
We also provide results for a synchronous-update threshold model version in the Appendix~\ref{sec:AIII}. 

We considered the threshold process both on regular and random networks. In the former case, we investigated mainly the two-dimensional square lattice, but also the triangular lattice and the three-dimensional cubic lattice were studied. Concerning random networks, we studied two-dimensional Voronoi graphs as well as (infinite-dimensional) random regular networks (RRNs).  
In the main text, we focus on one representative of these cases, the  threshold process on the square lattice, on the Voronoi graph, and on the random regular networks, while other structures and complementary arguments are discussed in the Appendices.

\section{Square lattice}
\label{sec:square}

\subsection{Numerical results}

We start the discussion of the steady state and relaxation dynamics of the model by presenting results of numerical simulations obtained for the two-dimensional square lattice with $\Theta=1/2$. 
The linear system size used in the simulations was $L=10^3$, and periodic boundary conditions were assumed in both directions. The system was initialized in a random state in which sites are activated independently with a probability $\rho_0$ and we followed the evolution of the density of active sites $\rho(t)$ up to time $10^6$. For each value of $\rho_0$ and $\lambda$, $10^3$ to $10^4$ independent runs, more for lower densities, were performed and an average of the time-dependent density was calculated. 

\begin{figure}[ht]
    \centering
    \includegraphics[width=1\linewidth]{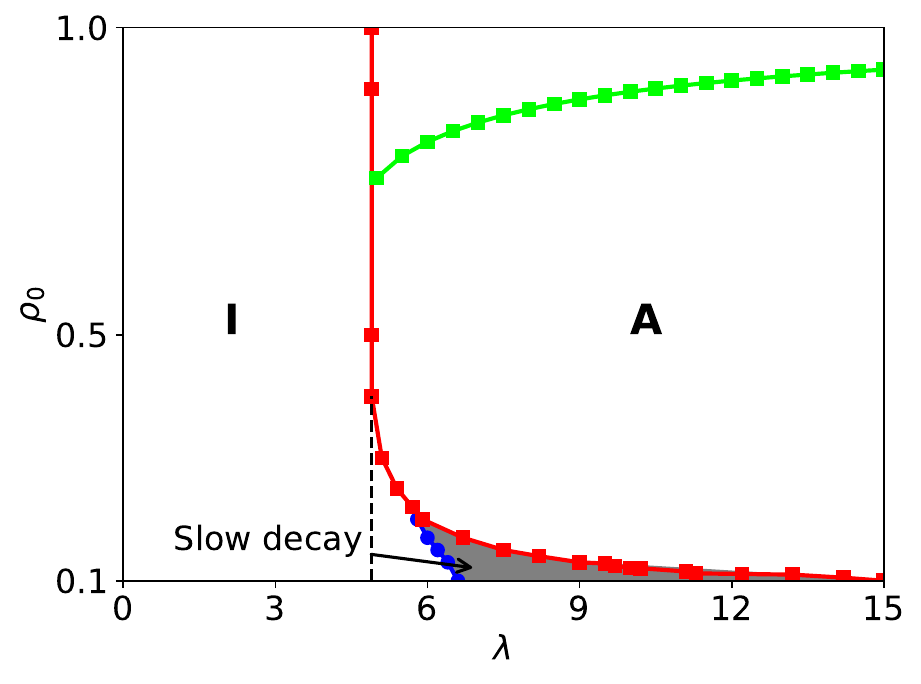}
    \caption{Numerically determined phase diagram of the threshold model on the two-dimensional square lattice with threshold $\Theta=1/2$. The phase boundary between the active and inactive phase is marked by red squares (measurement points). The red solid curve is a guide to the eye. In the gray-shaded area, the density is found to decrease algebraically in time. The black dashed line marks the transition point $\lambda_c=4.9$. The solid line with blue solid circles marks the lower bound of the slow decay region. The vertical coordinate of the green solid line with square markers gives the stationary density $\rho_\infty$ in region A. 
    \label{fig_pd} }
\end{figure}

First, we carried out simulations for a series of parameters $\lambda$ and $\rho_0$, and observed whether the density tends to zero or to a positive stationary value, the phase diagram of the model was constructed. 
As can be seen in Fig. \ref{fig_pd}, the phase diagram is qualitatively similar to that obtained by the mean-field approximation (see Fig. \ref{fig_mf}) but the critical value $\lambda_c=4.9$ marking the onset of the bistability region is considerably greater than the mean-field value. Here, the stationary density in the active phase, showing a discontinuous jump at $\lambda_c$, is also presented.

\begin{figure}[ht]
    \centering
    \includegraphics[width=1\linewidth]{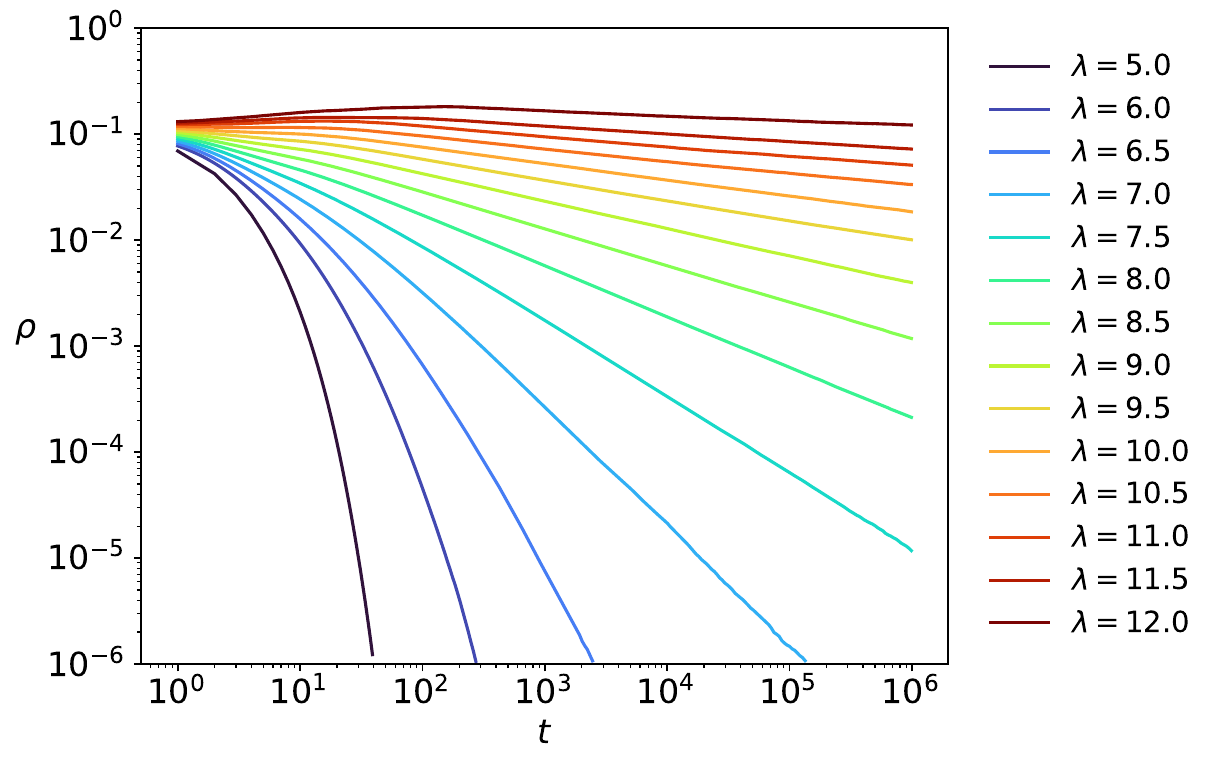}
    \caption{Time-dependence of the density decay, obtained by numerical simulations of the threshold model with $\Theta=1/2$ on a square lattice for several values of the activation rate $\lambda$. The initial density was $\rho_0=0.1$.}
    \label{continuous-time-model-decay}
\end{figure}

As opposed to the structure of the stationary phase diagram, the relaxation dynamics, i.e. the rate of tending to the steady state, displays a striking difference from the mean-field behavior. 
Although in most of the phase diagram the approach of the density to its stationary value is exponential in time, just like in the mean-field theory (except of the approach to $\rho_c$), an extended domain appears in the inactive phase in which the density is found to decay to zero algebraically as 
\be 
\rho(t)\sim t^{-\delta}
\ee 
In this slow-decay domain, which is bounded from above by the phase boundary line and is shaded in gray in Fig. \ref{fig_pd}, the decay exponent $\delta$ is found to vary (decrease monotonically) both with $\lambda$ and $\rho_0$.  
This non-universal relaxation is illustrated for $\rho_0=0.1$ and a series of $\lambda$ values in Fig. \ref{continuous-time-model-decay}.

\begin{figure}[ht]
    \centering
    \includegraphics[width=1\linewidth]{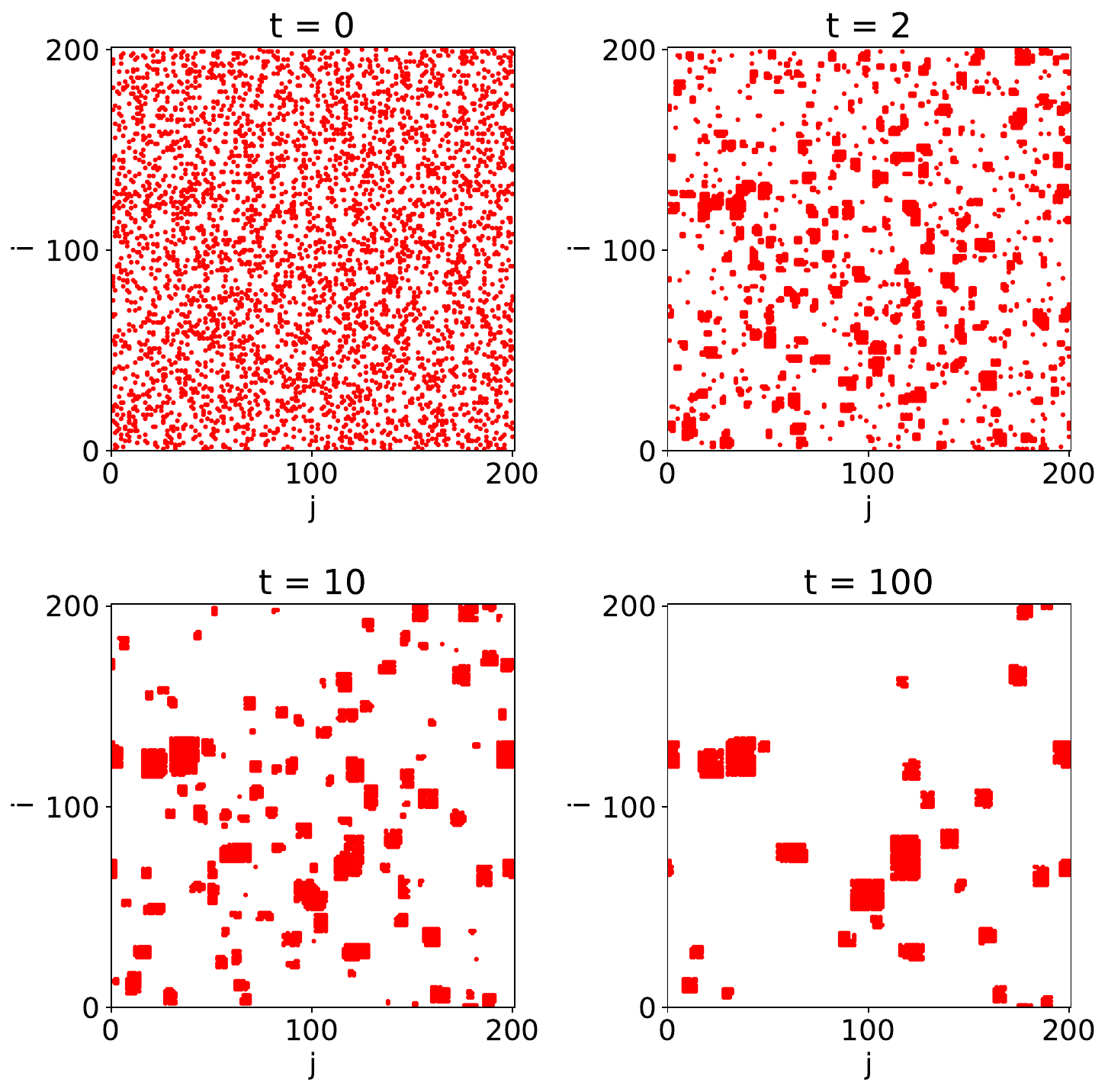}
    \caption{Snapshots of the configuration at different times, obtained for the model with $\Theta=1/2$, $\rho_0=0.1$, and $\lambda = 9$, on a square lattice of size $L =200$.}
    \label{fig:snaphots}
\end{figure}

The mechanism of this slow relaxation is illustrated by Fig. \ref{fig:snaphots}, which shows snapshots of the configuration of the system at different times. 
As can be seen in the figure, on a relatively short time scale, the system evolves to a configuration which is composed of isolated high-density clusters of roughly rectangular shape. Beyond this time scale, the relevant process of the dynamics is the gradual shrinking of clusters formed predominantly in the initial period, and ultimately their vanishing. 
The behavior in the latter period is a simple consequence of the threshold condition. Considering a completely active rectangular domain whose sides are parallel to the axis of the square lattice, and which is embedded in an otherwise empty lattice, it is easy to see that the activity cannot spread across the boundaries of the rectangle. The reason for this is that all the inactive external sites next to the rectangle have only one active neighbor, which is insufficient to get activated under the threshold condition with $\Theta=1/2$.   
This means formally that the threshold model on the square lattice with $\Theta=1/2$ is a highly reducible stochastic process i.e., the state space of the infinite system contains an infinite number of subspaces that cannot be left by the dynamics \cite{vankampen}.  
At late times, well beyond the period of cluster formation, the system is thus essentially composed of a set of non-communicating subsystems. 
As $\lambda>\lambda_c$, these subsystems are locally in the active phase and therefore get to the (inactive) absorbing state after a long time, resulting in the observed slow decay of the global density.

\subsection{A phenomenological model of slow decay}\label{sec:pheno}

In this section, we will investigate the phenomenon of slow decay described in the previous section in the frame of a simplified phenomenological theory.

We have seen in the simulations that, after a short initial period, the configuration of the system in the slow-decay phase essentially decomposes into a set of non-interacting clusters of different sizes. Similarly to the rare-region theory of the Griffiths phase,
the first question to clarify is how the mean time a given cluster needs to reach the local absorbing (empty) state depends on its size. 
For the standard contact process in its active phase, 
$\lambda>\lambda_c$, it is well known that a system with $N$ sites can get to the absorbing state by a rare fluctuation during which all sites become inactive. 
The rate of such an event is therefore exponentially low in $N$, and the corresponding mean-time-to-absorption is $\tau\sim e^{cN}$, where $c$ denotes a $\lambda$-dependent constant. 
In the threshold model, however, another, more efficient route to the empty state opens. 
As remarked earlier, activity cannot step over the boundaries of an active rectangular domain. Therefore, if a rare fluctuation erases a complete row or column at the boundaries of the rectangle, then this event is irreversible i.e., the erased row or column can never be rebuilt. Thus, active clusters will ratchet down (shrink) to the empty state gradually, by losing their actually outermost layer step by step.

\begin{figure}[ht]
    \centering
    \includegraphics[width=1\linewidth]{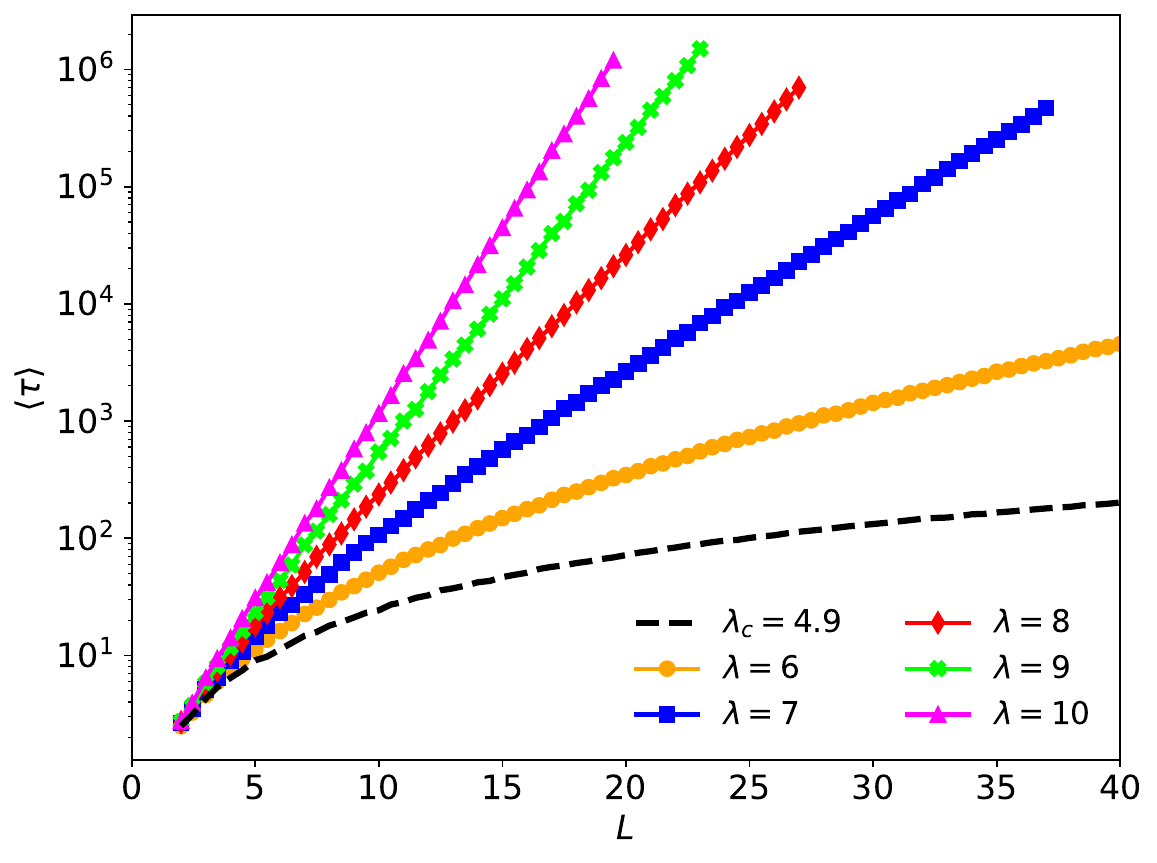}
    \caption{Mean time to absorption of fully active square-shaped domains as the function of linear size $L$ for different values of $\lambda$. Data were obtained by numerical simulations with a threshold parameter $\Theta=1/2$.     
    \label{fig:lifetimes-sq}
    }
\end{figure}

The mean time to irreversibly lose a boundary row or column with $l$ sites by a catastrophic fluctuation of deactivation events at each site is $\sim e^{cl}$. 
A consequence of this size-dependence is that the shorter-side layer of a rectangle will vanish on average much faster than the longer-side one; therefore, the aspect ratios of rectangles formed in the initial period will tend to one, i.e., they approach square shapes during the late-time (shrinking) period.   
As the aspect ratios are not much different from one even initially, the clusters can approximately be regarded as squares. 
The mean-time-to-absorption of a square-shaped cluster of size $l_0$, 
as follows from the above ratchet mechanism, is proportional to 
$\sum_{l=1}^{l_0}e^{cl}$, which sums up to a constant times $e^{cl_0}$. 
Thus, we obtain that, as opposed to the standard contact process, the mean-time-to-absorption of a cluster in the threshold model is exponential in the linear size rather than the volume:
\be 
\tau\sim e^{cl_0}.
\ee
This exponential dependence on the linear size is in agreement with simulation results on the mean time to absorption of square-shaped clusters in the regime $\lambda>\lambda_c$, as shown in Fig.~\ref{fig:lifetimes-sq}.

According to the mechanism outlined above, the shrinking of a cluster with an initial linear size $l_0$ can be modeled by a unidirectional jump process on the integers $0,1,2,\dots,l_0$ with jump rates $w_l=w_0e^{-cl}$ from state $l$ to state $l-1$, where $w_0$ denotes a constant. 
The mean time to reach state $l$ by this jump process is
\be
t_l=\frac{1}{a}(e^{cl_0}-e^{cl})
\label{tl}
\ee
with $a=w_0\frac{e^c-1}{e^c}$.
To simplify the treatment of the shrinking process, we will regard $l$ as continuous variable and we will ignore the stochasticity of the jump process by assuming a deterministic variation of $l$ with time\footnote{To support the validity of the deterministic treatment, we note that a stochastic treatment of the cluster deactivation process is not essential even in the phenomenological theory of the Griffiths phase in the disordered contact process. Indeed, assuming that each cluster vanishes deterministically after its mean lifetime has elapsed results in the same power-law decay of the global density as obtained by the stochastic treatment, as given in Eq. (\ref{tl})}.
This yields the time dependence of the linear size:
\be
l(t)=\frac{1}{c}\ln\left(e^{cl_0}-at\right)
\label{lt}
\ee
if $t<\frac{1}{a}\left(e^{cl_0}-1\right)$, otherwise $l(t)=0$.

Now, we turn to the question of how the distribution of the linear size of clusters changes in time. Let us denote the number of clusters having a linear size greater than $l$ per unit volume at time $t$ by $N_>(l;t)$,  and the initial distribution at $t=0$ by $N_>(l_0;0)$. 
Then, due to the deterministic nature of Eq.~\eqref{lt}, one can convince oneself that, for a fixed time $t\ge 0$, 
\be
N_>(l(t);t)\equiv N_>(l_0;0).
\label{Nt}
\ee
We note that the above deterministic treatment of the shrinking process leading to Eq. (\ref{Nt}) is equivalent to describing the original jump process by a Fokker-Planck equation with the omission of the diffusive term \cite{vankampen}. 

The normalized probability density that a randomly chosen cluster has  linear size $l$ is obtained as
\be
p(l;t)=\frac{1}{N_>(0;t)}\left|\frac{dN_>(l;t)}{dl}\right|.
\label{rho_gen}
\ee

\begin{figure}[ht]
    \centering
    \includegraphics[width=1\linewidth]{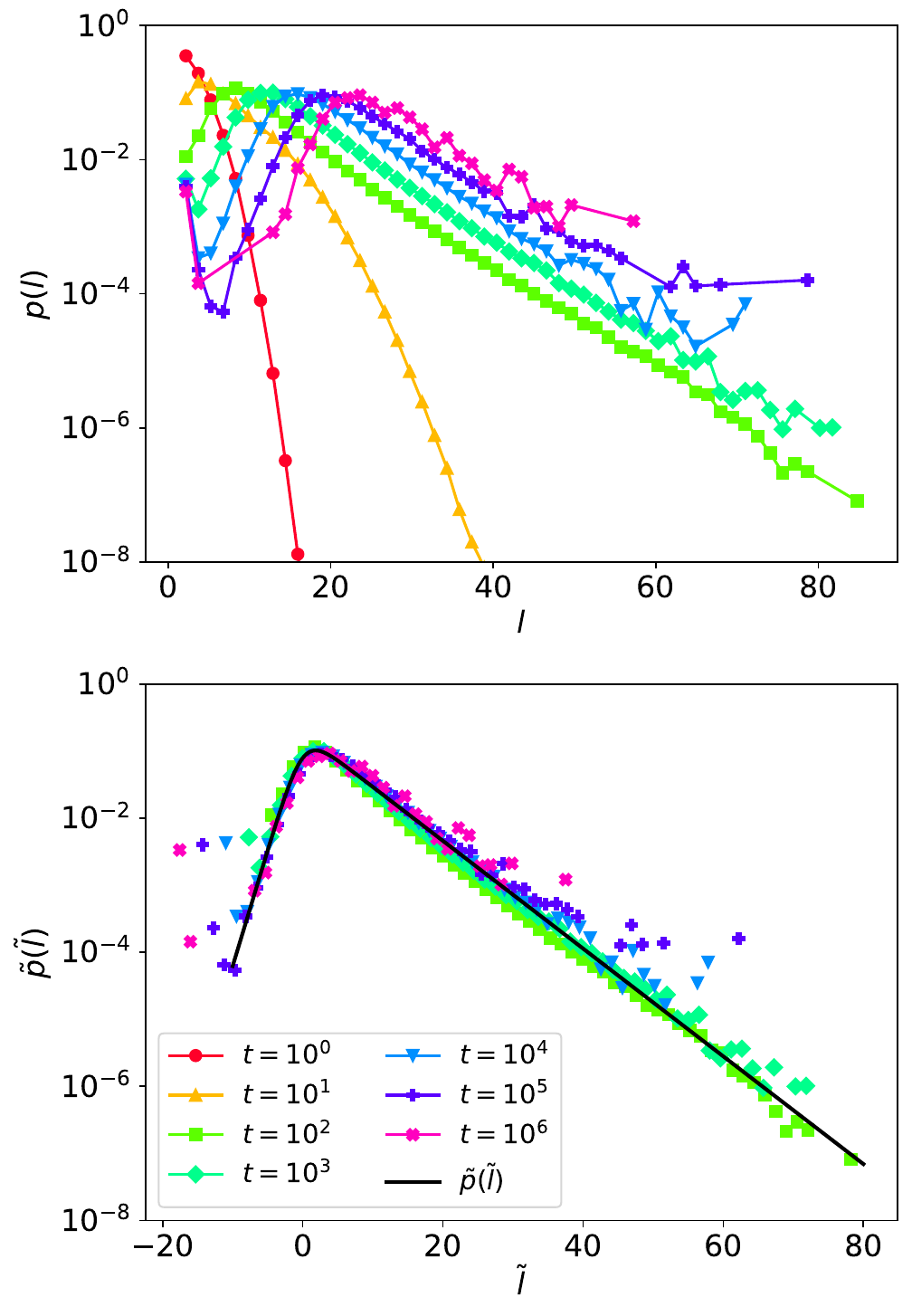}
    \caption{ Top. Linear size distribution obtained from simulations
    on the square lattice with $L=200$ at different times. The simulations were started with an initial density $\rho_0$, and parameters $\Theta=1/2$ and $\lambda = 9$ were used. The number of samples was at least $25000$ ($10^7$ for shorter times).     
    Bottom. Distributions of the scaling variable $\tilde l=l-\frac{1}{c}\ln(at)$ for $t\geq10^2$, using $a=1$.
    The solid black curve is the scaling function given in Eq. (\ref{rho_l_tilde}) with $b=0.185$ and $c = 0.8$. 
    \label{fig_ldist_square}
    }
\end{figure}

To determine the initial distribution of cluster size $l$, we resorted to numerical simulations. Taking snapshots of the configuration of the system at different times, we identified activity clusters as percolation clusters and estimated their linear size by taking the square root of the number of the constituting active sites.   
Distributions of the linear size obtained in this way at different times are plotted in Fig. \ref{fig_ldist_square}.  
Note that the size distribution $N_>(l_0;0)$ that is the initial distribution of the phenomenological model is identified with the distribution in the original process at the end of the initial cluster formation period, at which the large-$l$ tail of the distribution develops. 
In Fig. \ref{fig_ldist_square}, we can see that until $t\sim 10^2$ an exponential tail of the distribution emerges, and at later times, the distribution is shifting toward higher $l$ values but its shape remains essentially unchanged. 

Using an exponential initial distribution, $N_>(l_0;0)=Ce^{-bl_0}$ with constants $C$ and $b$ as an input for the deterministic theory, we obtain through Eqs. (\ref{Nt}) and (\ref{rho_gen}) the following time-dependent probability density of linear size:  
\be
p(l;t)=\frac{b(at+1)^{b/c}}{(ate^{-cl}+1)^{b/c+1}}e^{-bl}.
\label{rho_l_t}
\ee
If $t\gg 1$, it has a scaling form 
\be
\tilde p(\tilde l)
=\frac{be^{-b\tilde l}}{(e^{-c\tilde l}+1)^{b/c+1}}
\label{rho_l_tilde}
\ee
in terms of the scaling variable 
\be
\tilde l=l-\frac{1}{c}\ln(at).
\label{ltilde}
\ee
As can be seen in Figure~\ref{fig_ldist_square}, the numerically obtained distributions show a data collapse in terms of the scaling variable $\tilde l$ at late times, and also the scaling function obtained by the deterministic theory fits well to the numerical distributions.

Next, we turn to calculating the time dependence of the global density within the deterministic theory. 
For this, we may write 
\be 
\rho(t)=\int_0^{\infty}\left|\frac{dN_>(l;t)}{dl}\right|l^2dl.
\label{rho_int}
\ee
Integrating by parts, this can be rewritten as 
$\rho(t)=2\int_0^{\infty}N_>(l;t)ldl$, which valid is valid if $N_>(l;t)$ vanishes faster than $1/l^2$. 
For an exponential initial distribution, we obtain by Eq. (\ref{Nt}),
\be
N_>(l;t)=C[e^{cl}+at]^{-b/c}.
\label{Nexp}
\ee
Then the expression for the density takes the form
\be
\rho(t)=2C(at)^{-b/c}\int_0^{\infty}[1+e^{cl}/(at)]^{-b/c}ldl.
\ee
The first factor of the integrand cuts off exponentially if $e^{cl}/(at)\gg 1$, thus we may approximate the integral by 
$\int_0^{l^*}ldl$ with $l^*=\frac{1}{c}\ln(at)$. 
This leads ultimately to an algebraic decrease of the density with a multiplicative logarithmic correction:
\be 
\rho(t)\sim t^{-b/c}\ln^2(at).
\label{rhot}
\ee
The decay exponent $\delta$ observed in simulations is thus related to the parameters of the phenomenological theory via $\delta=b/c$. 

It is important to note that the above derivation, presented for two-dimensional regular lattices, can be generalized to
other, higher dimensional graphs, by expressing $\rho(t)$ via the cluster-size distribution and the relationship between lifetime and size. 
Indeed, we have also checked the density decay behavior for two-dimensional triangular and for higher dimensional
hypercubic lattices and found similar behavior when the corresponding  threshold $\Theta$ caused a discontinuous
transition. In particular, for the triangular lattice with $\Theta=1/2$, corresponding to three-neighbor activation, we obtained power-law decay. We also show slow decay in $d=3$ and $d=4$ for $\Theta=3/6, 3/8$, respectively, without going into deeper analysis in Appendix~\ref{app:higher}. Furthermore, we also present numerical results obtained by a synchronous-update version of the model in Appendix~\ref{sec:AIII}. 

\subsection{Stability of the slow density phase}\label{sec:stab}

Next, we turn to the question of stability of the slow-decay phase. To illustrate the problem, let us first consider the special case $\mu=0$ of the model, with the process started from an initial density $\rho_0$ as before. This process is equivalent to bootstrap percolation~\cite{bootstrap}. Here, a square-shaped domain of active sites with a size at least $1/\rho_0$ is known to grow indefinitely  and make the whole system active with a finite probability. Indeed, for a large active domain embedded in a system where activity is randomly deposited, there can be active sites adjacent to the boundary of the domain, triggering the formation of a new complete row of active sites. The resulting enlarged domain then possesses a new boundary where the same mechanism can recur, leading to successive domain growth. 
Since active squares of arbitrary size occur initially with a finite probability, the infinite  system will be ultimately active for any small $\rho_0>0$. For finite systems of size $L$, however, there is a characteristic initial density $\rho_0\sim 1/\ln L$, below (above) which the bootstrap percolation process stops in a configuration with $\rho<1$ ($\rho=1$) \cite{holroyd}. 
 
\begin{figure}[ht]
    \centering
    \includegraphics[width=1\linewidth]{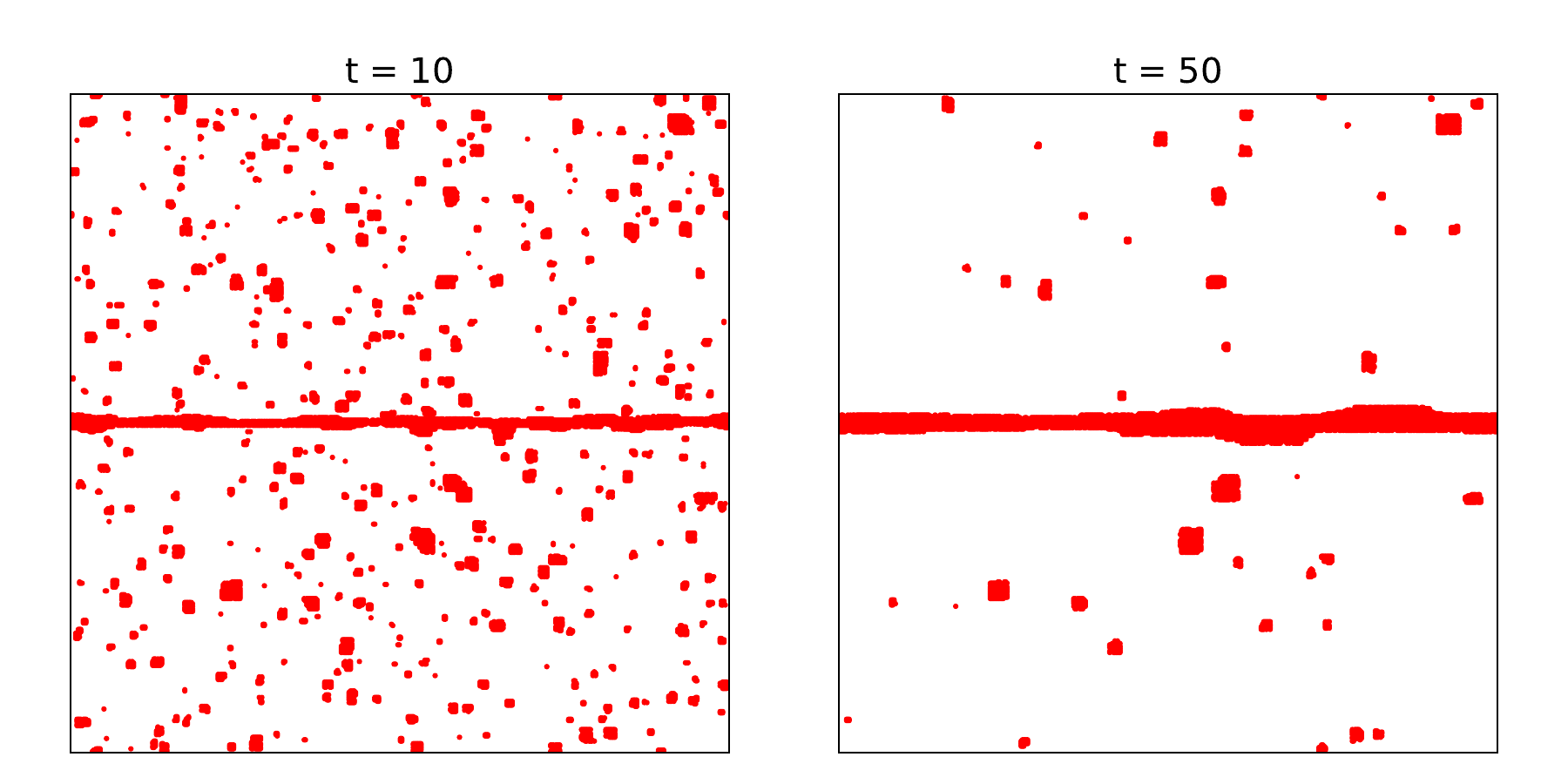}
    \caption{Core cluster growth at two different times for a square lattice with $L=512$, $\rho_0=0.1$, $\lambda=7.5$, $\Theta=2/4$ with an active ring condition.}
    \label{fig:core-cluster}
\end{figure}

For the model with $\mu>0$, the same question arises: whether an active domain of size greater than $1/\rho_0$ can grow indefinitely, thereby destabilizing the slow decay of density and turning the steady state into a finite-density state rather than the absorbing one. 
In the threshold process, a rectangular cluster with sides parallel to the axes is able to grow by other clusters in its boundary zone (composed of the nearest and next-to-nearest external layers) , which induce the addition of a whole stripe to the growing cluster. 
What is different compared to bootstrap percolation ($\mu=0$) is that there is a steady loss of other (small) clusters which promote growth, making the conditions of growth more and more unfavorable.  
At the same time, the larger the size of the cluster is, the more favorable is its further growth. Thus, there is a competition between these two circumstances, and we have to compare the rates of these competing processes to decide whether indefinite growth is possible or not. 
We will examine this question within the phenomenological theory introduced in section \ref{sec:pheno}.  
According to this, the number density of clusters $n(t)\equiv N_>(0;t)$, as it follows from Eq. (\ref{Nexp}), decreases as 
\be 
n(t)\simeq n_0t^{-\delta}
\label{nt}
\ee
at late times, with a constant $n_0$ and $\delta=b/c$ 
Furthermore, as can be concluded from Eq. (\ref{ltilde}), 
the typical size $\xi(t)$ of clusters present in the system increases with time asymptotically as 
\be 
\xi(t)\sim \frac{1}{c}\ln(at).
\label{xit}
\ee
Let us assume that we have an infinitely large lattice and an infinite line of sites (parallel to an axis of the lattice) which are kept active (by setting $\mu=0$ there). In other words, for the two half-planes on the two sides of this line, we have a so-called active-wall boundary condition. As before, other sites of the system are activated initially with a probability $\rho_0$. Then, as time elapses, the active cluster starting out from the active line 
(that we will call core cluster) is steadily growing and becoming thicker and thicker while the number of other clusters decreases as shown by the snapshots in Fig.~\ref{fig:core-cluster}. 
We are interested in the dependence of its mean width $l(t)$ on time. 
According to the considerations presented in Appendix \ref{app:growth}, 
we obtain that the increase of the mean width depends on the decay exponent $\delta$ as follows:
\be
l(t)\sim\begin{cases}
 t^{1-\delta}\ln t & {\rm if} \quad \delta<1 \\
 (\ln t)^2 & {\rm if} \quad \delta =1  \\
 \ln t & {\rm if} \quad \delta>1.  
\end{cases}
\label{ltcases}
\ee

\begin{figure}[ht]
    \centering
    \includegraphics[width=1\linewidth]{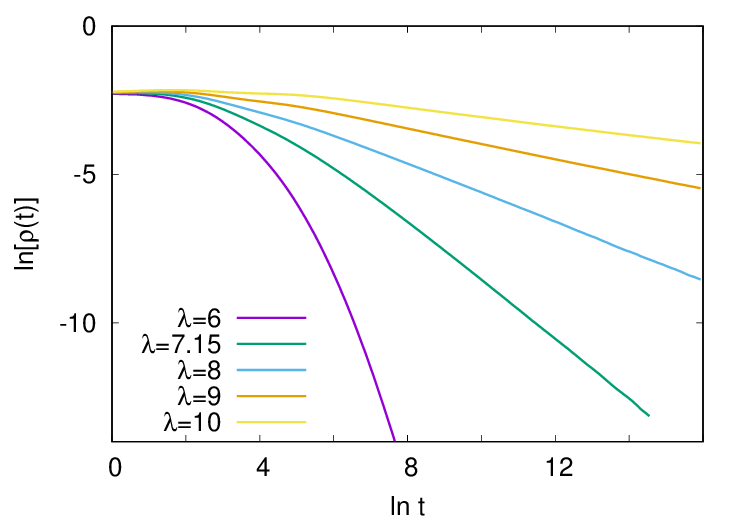}
    \caption{Time dependence of the density for the square lattice with $\Theta=1/2$, $\rho_0=0.1$ and different values of $\lambda$.  The linear system size was $L=10^4$.
    \label{fig.1.eps}  }
\end{figure}

\begin{figure}[ht]
    \centering
    \includegraphics[width=1\linewidth]{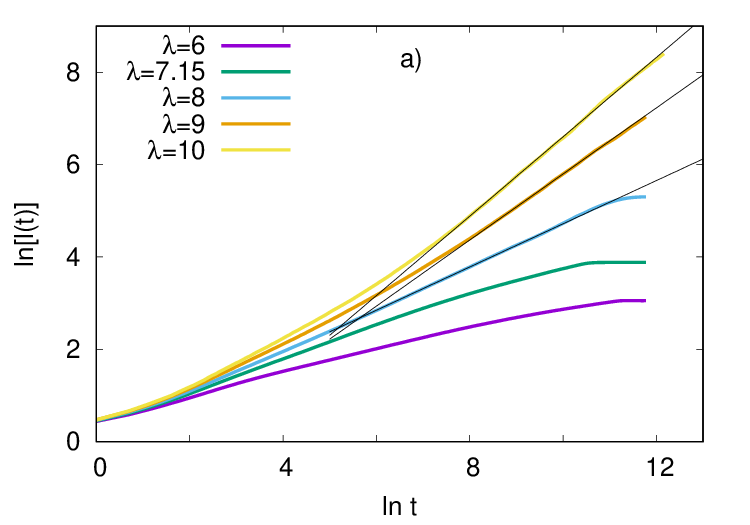}
     \includegraphics[width=1\linewidth]{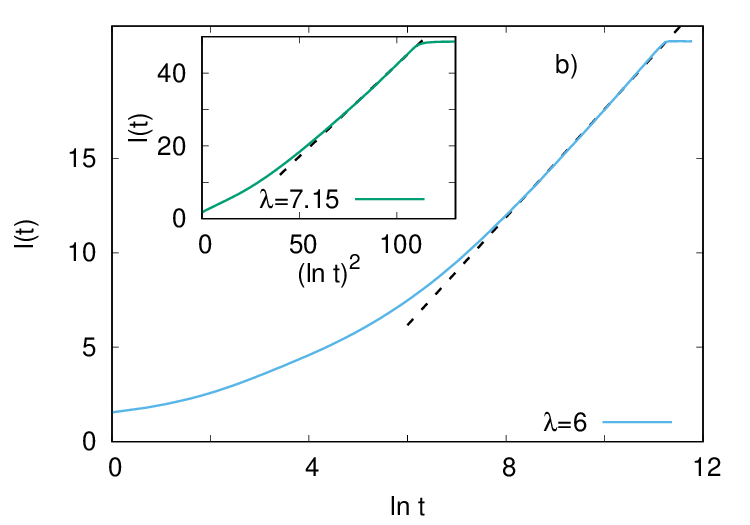}
    \caption{a) Time dependence of the mean width of the core cluster for random initial state with $\rho_0=0.1$ and active-ring boundary condition, for different values of $\lambda$ on a log-log plot.
    b) Linearized plots of the data in the regime $\delta\ge 1$. The straight lines are linear fits to the data with slopes $\kappa$. 
    \label{fig_line} }
\end{figure}

In order to check the dependence of the mean width of the core cluster on time, we first estimated the decay exponent $\delta$ for a random initial condition 
with $\rho_0=0.1$ and different values of $\lambda$ in the slow-decay phase. We also determined the control parameter at which $\delta=1$ to be $\lambda=7.15(2)$. The data used for the estimation of $\delta$ are plotted in Fig. \ref{fig.1.eps}, while the estimates can be found in Table \ref{table_exp}. 
Then we performed numerical simulations on a square lattice (torus) of linear size $L=10^4$ in the presence of a steadily active ring (of size $L$) along an axis of the lattice. Other sites were activated initially with a probability $\rho_0=0.1$. We measured the mean width $l(t)$ of the core cluster, which is plotted against time in Fig. \ref{fig_line} for different values of $\lambda$.  
As can be seen, $l$ increases slower-than-algebraically for $\delta\ge 1$, while for $\delta<1$, we find an algebraic increase as $l(t)\sim t^{\kappa}$, in agreement with the predictions of the phenomenological theory. In the latter regime, Table \ref{table_exp} shows the estimated exponents $\kappa$. The relation $\kappa=1-\delta$, bearing in mind that there are also multiplicative logarithmic corrections in both $\rho(t)$ and $l(t)$, is satisfactorily fulfilled.
\begin{table}[ht]
\begin{center}
\begin{tabular}{|c|c|c|}
\hline $\lambda$ & $\delta$ & $\kappa$   \\
\hline
\hline 7.15 & 1.00(1)  & - \\
\hline 8 & 0.50(1)  & 0.47(1)  \\
\hline 9 &  0.25(1) & 0.72(2)  \\
\hline 10 &  0.15(1) & 0.86(1)  \\
\hline
\end{tabular}
\end{center}
\caption{\label{table_exp} Numerical estimates of the decay exponent $\delta$ and the exponent $\kappa$ characterizing the increase of mean width of the core cluster (see text) for $\rho_0=0.1$ and different values of $\lambda$. } 
\end{table}

Now, let us return to the original setup in which there is no active wall in the system. Instead of this, let us assume that there is initially an active square of size $l_0\gg 1/n_0$ in the system. The probability of this is exponentially small in $l_0$, nevertheless, this probability is finite for finite $l_0$, and in a sufficiently large system such an active square will form with an $O(1)$ probability. 
The condition $l_0\gg 1/n_0$ guarantees that, initially, there will be activity clusters in the boundary zone of the big cluster with a high probability, so that it can start to grow.
As the dynamics of the threshold process starts, the linear size of the square will increase according to the functions $l(t)$ obtained above. But, now the side lengths of the square are finite (initially $l_0$), and we have to check whether the condition
\be 
l(t)\gg 1/n(t)
\label{cond}
\ee
holds also for later times, which is a necessary condition for an indefinite growth. In words, this condition means that there are many touching clusters also at a later time $t$.
We know that $1/n(t)\sim t^{\delta}$, and it is evident that the condition in Eq. (\ref{cond}) cannot be fulfilled for $\delta>1$. In the domain $\delta<1$, we obtain by comparing Eqs.~(\ref{nt}) and (\ref{ltcases}) that the condition is fulfilled if
\be\label{eq:half}
\delta\le \frac{1}{2}.
\ee
At $\delta=\frac{1}{2}$ the condition is fulfilled but just barely (logarithmically).
This implies that, in the regime $\delta\le \frac{1}{2}$, the low-density state  must be  unstable, i.e. if a sufficiently large active domain appears (which occurs with a high probability for large enough system sizes) the state will turn to a high-density state. In other words, in an infinite system initialized with uncorrelated probabilities $\rho_0$, the asymptotic slow-decay behavior exists only for $\delta>1/2$.
The above considerations suggest that the density decay with $\delta\le 1/2$ observed in simulations must be a transient behavior and a crossover must occur to an increase in an infinite system.

\section{Voronoi–Delaunay network}

Up to now we investigated the slow decay of density on regular lattices.
Now we test if this phenomenon can also be seen in heterogeneous networks.
Spatial connections are established through the Delaunay triangulation, the dual structure of the Voronoi diagram~\cite{Voronoi1908}. The Voronoi tessellation partitions the plane according to the nearest-neighbor criterion, while the Delaunay triangulation connects pairs of points whose Voronoi cells share a common edge~\cite{Delaunay1934}. The resulting planar network that captures local spatial relationships \cite{Barthlemy2011} will be used as the underlying graph used in the subsequent analyses.
Unlike regular lattices, Voronoi-Delaunay networks exhibit structural heterogeneity arising from the random spatial distribution of nodes. As a consequence, vertices may possess different numbers of neighbors, leading to a nontrivial degree distribution centered around an average degree $\langle k \rangle = 6$, while preserving the finite-dimensional nature of the underlying space. This feature makes such networks particularly suitable for investigating the effects of quenched spatial heterogeneity on dynamical processes in finite-dimensional systems \cite{DeOliveira2008a,Barghathi2014,DeOliveira2016}.

\begin{figure}[ht]
    \centering
    \includegraphics[width=1\linewidth]{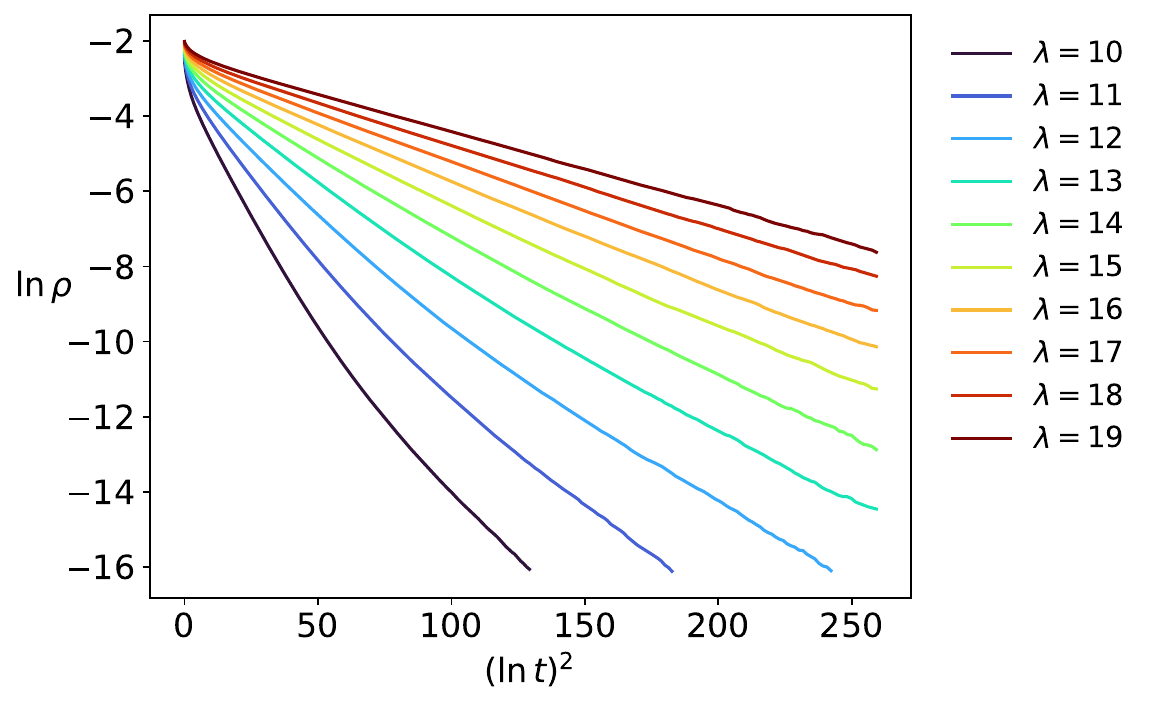}
    \caption{Density decay on a 2D Voronoi–Delaunay network with $N=10^7$ sites, $\rho(0)=0.12$ and $\Theta= 0.4$. Single graph realization. }
    \label{fig:voronoi0devay}
\end{figure}

Similar to regular lattices, we found a region in which slowly shrinking isolated activity clusters emerge at late times. However, in spite of the qualitative similarity, the dynamics shows quantitative differences compared to regular lattices.  
As we can see on Fig.~\ref{fig:voronoi0devay} in case of a single realization on a large graph
with $N=10^7$ sites, the density decays slower than exponentially but faster than algebraically. The functional form is found to be more or less compatible with an enhanced power law:
\be 
\rho(t)\sim e^{-C(\ln t)^2},
\ee
where $C$ denotes a non-universal constant.

\begin{figure}[ht]
    \centering
    \includegraphics[width=1\linewidth]{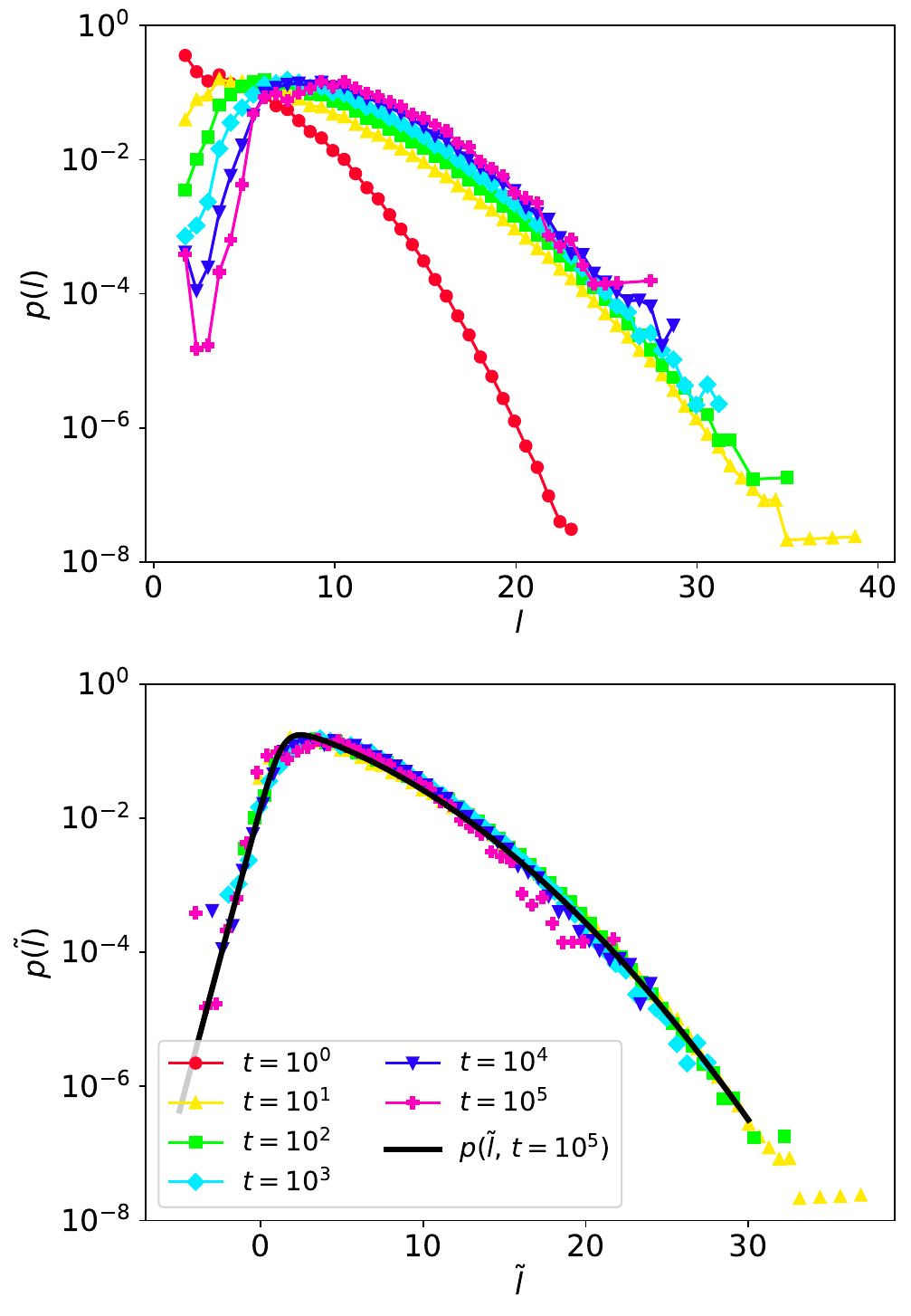}
    \caption{Top. Linear size distribution obtained from simulations on Voronoi–Delaunay network for $t=10^0$ ($10^7$ samples), $t=10^1$ (4928400 samples), $t=10^2$ (1126790 samples), $t=10^3$ (168060 samples), $t=10^4$ (50260 samples) and $t=10^5$ (15650 samples). 
    Bottom. Distribution of the scaling variable in Eq. (\ref{ltilde2}), for $t\geq10$. The solid line is the theoretical prediction in Eq. (\ref{rho_gaussian}). The results are obtained for a Voronoi graph with $N=40000$ sites, $\rho(0)=0.12$, $\lambda = 18$, $\Theta=0.4$, $a = 25$, $b=0.011$ and $c = 2.1$.  
    \label{fig_voronoi_dist}
    }
\end{figure}

We also studied the distributions of linear cluster size at different times, in the same way as for the square lattice.  
As opposed to regular lattices, we found the tail of the distribution at short times (after the period of cluster formation), to decrease according to $p(l)\sim e^{-bl^2}$, where $b$ denotes a non-universal constant, see Fig \ref{fig_voronoi_dist}.
Applying the phenomenological theory presented in the previous section to the case of a Gaussian initial distribution
\be
p(l_0;0)=2\sqrt{\frac{b}{\pi}}e^{-bl_0^2},
\ee
we obtain for the distribution of cluster sizes at time $t$:
\be
p(l;t)
=\frac{2\sqrt{\frac{b}{\pi}}e^{-\frac{b}{c^2}\ln^2(at+e^{cl})}\frac{1}{ate^{-cl}+1}}{1-{\rm erf}[\frac{\sqrt{b}}{c}\ln(at+1)]}.
\label{rho_gaussian}
\ee
The maximum position of this distribution increases at late times as 
$l^*(t)\simeq \frac{1}{c}\ln\frac{at}{\ln(at)}-\frac{1}{c}\ln(2b/c^2)$
and, accordingly, the variable 
\be 
\tilde l=l-\frac{1}{c}\ln\left[\frac{at}{\ln(at)}\right]
\label{ltilde2}
\ee
has a limit distribution 
\be 
\tilde p(\tilde l)=\frac{2b}{c}e^{c\tilde l-\frac{2b}{c^2}e^{c\tilde l}}
\ee
when $t\to\infty$, which is known as the Gumbel distribution. Nevertheless, the convergence to this limit distribution is extremely slow, being satisfactory if $\tilde l\ll \frac{1}{c}\ln[\ln(at)]$. 
The theoretical distribution in Eq. (\ref{rho_gaussian}) fits well to the numerically obtained ones, as demonstrated in Fig.~\ref{fig_voronoi_dist}.
By using $N_>(l;t)={\rm erfc}[\frac{\sqrt{b}}{c}\ln(at+e^{cl})]$ and Eq. (\ref{rho_int}), we obtain that, at long times, the global density decays in leading order as 
\be 
\ln\rho(t)\sim -\frac{b}{c^2}\ln^2(at),
\ee
in agreement with the numerical results.
Thus the density decays in time faster than for regular lattices. One can show that unlike for a power-law decay, this enhanced power-law decay is always stable against the formation of arbitrarily large clusters.

\section{Random regular graphs}\label{sec:ARR}
In the framework introduced by Erdős and Rényi, all connections in a random regular network are chosen with equal probability~\cite{ErdosRenyi1959}, thus the node degree can take an arbitrary value. This introduces quenched randomness in the system. 
Random regular networks (RRNs) are less heterogeneous since the degrees of all nodes are kept to be the same, although the nodes are still interconnected randomly~\cite{Bollobas2001}.
An important property of RRNs is small-worldness and a corresponding infinite topological dimension \cite{BarabasiPosfai2016}.
For this, phase transitions of dynamical processes on such networks are expected to be described by mean-field theory.

\begin{figure}[ht]
     \centering
     \includegraphics[width=1\linewidth]{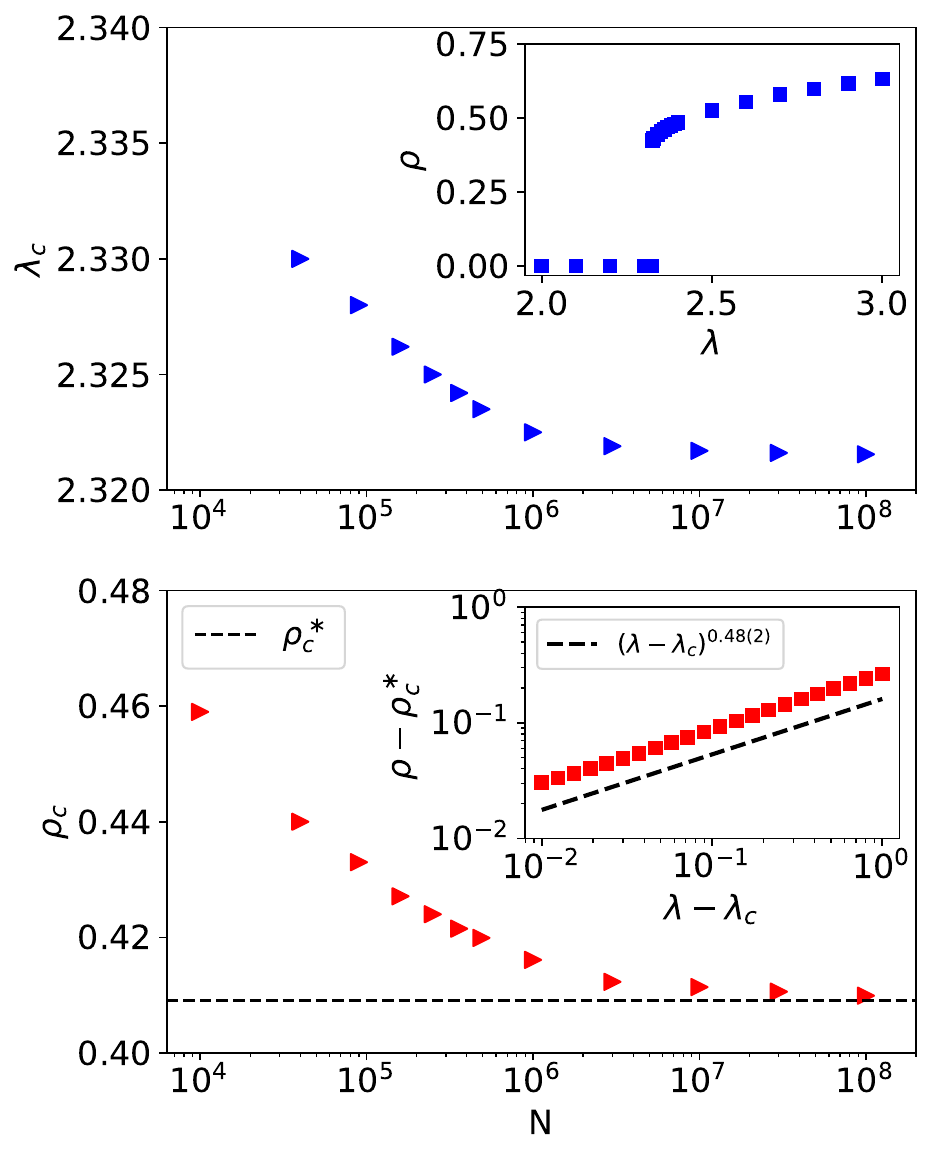}
     \caption{Finite-size scaling for the threshold model with $\Theta = 2/6$ on RRNs with degree $k=6$.
     Top. Finite-size scaling of $\lambda_c$.  The inset shows the steady-state density of active sites for $N=2.5\times10^5$ sites. The transition is estimated to be at $\lambda_c = 2.325$. Bottom. Finite-size scaling of the critical density of active sites $\rho_c$. The dashed horizontal line indicates the extrapolated value of critical density, $\rho_c^\ast \equiv \rho_c(N\to\infty,\,t\to\infty)=0.409(3)$.
     The inset demonstrates the algebraic approach of the stationary density to its limiting value with an exponent $\beta'=0.48(2)$.    \label{fig:RRdec2}
     }
\end{figure}

Besides investigating whether a slow-decay phase appears for the threshold process on RRNs, we also aimed at testing the validity of mean-field description at $\lambda_c$.
For this purpose, we constructed RRNs consisting of $N = 2.5\times 10^5$ nodes with a uniform degree $k=6$ and performed simulations of the  threshold contact process with two different values of the threshold parameter $\Theta$. 
First, we considered the case in which the activation of a site requires at least two active neighbors, corresponding to $\Theta = 2/6$. 
We started our analysis with the behavior of stationary density of active sites evolved from the initial density $\rho_0 = 1$ near $\lambda_c$.  
Here, small variations in the control parameter 
are found to lead to a discontinuous jump in the stationary density, separating the absorbing state from an active phase, as shown in Fig.~\ref{fig:RRdec2}, in agreement with mean-field theory.
In the vicinity of $\lambda_c$ we found strong finite-size effects, therefore we performed a finite-size scaling analysis to precisely estimate $\lambda_c$ and critical density $\rho_c$ in the thermodynamic  limit.
As can be seen in Fig.~\ref{fig:RRdec2}, as $\lambda$ tends to $\lambda_c$ from above, the stationary density $\rho$ approaches $\rho_c$ algebraically as 
\be
\rho - \rho_c \sim (\lambda - \lambda_c)^{\beta'}. 
\ee
The exponent appearing here is estimated to be $\beta'=0.48(2)$, which is compatible with the prediction $\beta'=1/2$ of the mean-field theory.

\begin{figure}[ht]
    \centering
    \includegraphics[width=1\linewidth]{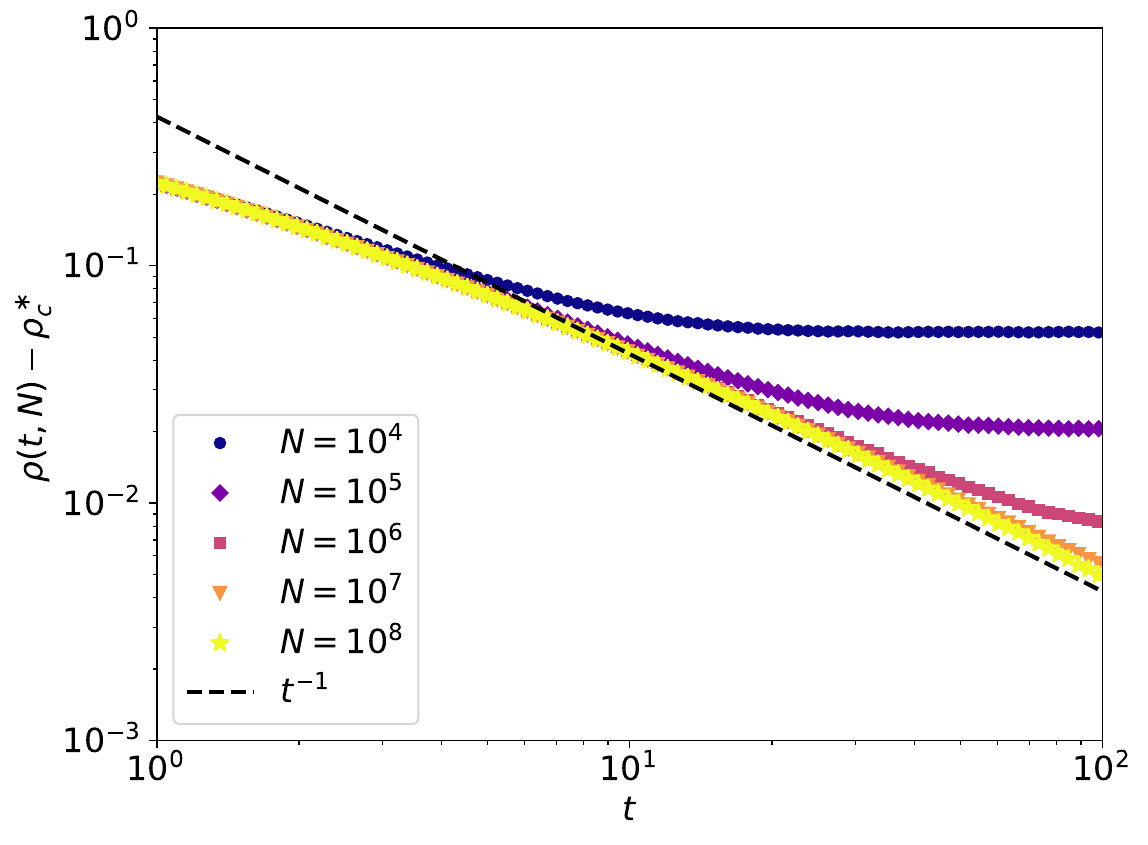}
    \caption{Time evolution of $\rho(t,N) - \rho_c^\ast$ for different system sizes $N$. The data suggest a scaling behavior consistent with $\rho(t) - \rho_c^\ast \sim t^{-1}$ as $N \to \infty$.     
    \label{fig:RR-scaling}
    }
\end{figure}

As shown in Fig. \ref{fig:RR-scaling},
the time-dependence of the density at $\lambda_c$, still with the initial density $\rho_0=1$, is found to follow the law 
 \be
 \rho(t) - \rho_c \sim t^{-1}
 \ee
 again in accordance with mean-field theory.

Thereafter, in search of a slow-decay phase, we performed simulations starting with a series of initial densities $\rho_0<1$. According our results, no signs of a slower-than-exponential decay showed up for any value of $\rho_0$ and $\lambda$. 

Finally, we investigated the model with a stricter threshold condition, namely with $\Theta = 3/6$. 
Similarly to the case $\Theta=2/6$, simulations with various values of $\rho_0<1$ and $\lambda$ do not show slow-decay phenomenon. 

\section{Discussion}

We have investigated in this work the relaxation dynamics of threshold contact processes in the two-phase coexistence region. 
By numerical simulations and a phenomenological theory, we found a general slower-than-exponential temporal decrease of the order parameter in an extended region of the parameter space.
The mechanism of this slow relaxation roots in the highly reducible nature of the process, which is a consequence of the threshold condition imposed on the activation rules. The configuration of the system evolved from a sufficiently low initial density will break up into spatially separated domains of activity. Owing to the threshold condition, these domains do not communicate with each other and, being locally in the active phase, reach the inactive state very slowly. 

Qualitatively, this behavior is reminiscent of the slow decay of density in the inactive Griffiths phase of the contact process with quenched random rates~\cite{Vojta2005}. In that model, spatial fluctuation of quenched disorder lead to the emergence of locally supercritical rare-regions immersed in a predominantly subcritical background. As activation across subcritical regions is of low rate, the activity stored in rare-regions  will vanish essentially independently, leading to a slow, algebraic decay of the global density in the inactive phase.  
Besides similarities to Griffiths phase of the quenched disordered contact process there are also differences. 
First, in the threshold contact process, the rare active regions are not triggered by quenched disorder, but in the absence of such, they form spontaneously and randomly at the early period of stochastic dynamics.  
Second, the communication between rare regions is perfectly quiescent in the threshold contact process, whereas for the quenched contact process independence is valid only effectively, compared to the lifetime of rare regions. 
Third, the activity in rare regions of the quenched contact process can only be erased by a rare fluctuation covering the whole region, and as such, leads to a lifetime that is exponential in the volume $l^d$ of the rare region. As opposed to this, in the threshold contact process, the deactivation occurs in a more efficient way, through consecutive and irreversible erasures of parts having a dimension $d'$ lower than the embedding space. This leads to a lifetime increasing exponentially in $l^{d'}$. For the example of the model with threshold parameter $\Theta=1/2$ on the square lattice, studied in most detail in this work, the deactivation of a rare-region of rectangular shape is realized by erasures of one-dimensional rows or columns. 

We found this kind of slow-decay mechanism to be present generally in regular lattices with appropriately chosen threshold parameter $\Theta$, irrespectively of the spatial dimension and lattice type. 
The quantitative dependence of global density on time is determined by two factors. First, the dependence of lifetime of an active domain on its linear size. Its asymptotic form follows from considering the shape of active domains for a given lattice type and threshold parameter. 
Second, it is also influenced by the tail of the size distribution of active clusters emerging after the initial period, which is determined again by the lattice type and the threshold parameter in a non-trivial way. For the size distribution, we have only results of numerical simulations at our disposal, and an analytical progress addressing this question would be highly desirable.  

For a power-law decrease of the global density, $\rho(t)\sim t^{-\delta}$, which is the case for a square lattice with $\Theta=1/2$, 
we pointed out by a phenomenological theory that the slow decay is stable and valid asymptotically only if $\delta>1/2$.
Otherwise, the decaying state must be unstable against an unbounded nucleation process around active clusters, the linear size of which is well beyond a critical nucleation radius $l_0\sim 1/\rho_0$. 
Not that this limitation of the stable slow-decay phase to the range $\delta>1/2$ is another difference from the decay in the Griffiths phase of the quenched contact process, where the exponent $\delta$ can be arbitrarily close to zero. 
Nevertheless, as the tail of initial cluster-size distribution decreases rapidly, such a nucleation process in the metastable region could be observed only in extremely large systems. For example, in the case of the square lattice of linear size $L$ and $\Theta=1/2$, the condition for nucleation is given by $\ln L\gg l_0$. In finite systems below this characteristic system size, the global density still follows a power-law decay with an effective exponent smaller than $1/2$.

We also pointed out that the regularity of the underlying lattice is not essential for the appearance of slow-decay phenomenon. This has been tested by studying the threshold model on random Voronoi-Delaunay networks, which, in spite of their structural heterogeneity, still have a finite topological dimension ($d=2$). According to our numerical results, the active regions have an irregular shape here, but they can still be spatially separated from each other, so that the mechanism of slow decay phenomenon holds to be valid. A quantitative difference from other two-dimensional lattices is that the tail of the linear cluster-size distribution is found to better fitted to a Gaussian form rather than to an exponential function. As a consequence, the decay of global density is found here to follow an enhanced power law. Nevertheless, a crossover of the tail to an exponential asymptotics, implying a power-law decay of density, may not be excluded.   

Further probing the necessary conditions of the slow-decay phenomenon, we relaxed the locality property of the underlying lattice by considering the threshold contact process on random regular networks. These networks are known to possess small-world characteristics, which hinder the formation of disjunct clusters. In other words, unlike for finite dimensional regular or random lattices, no smooth hypersurfaces can be constructed here across which the spreading of activity is blocked by the threshold condition. As a consequence, the mechanism of slow-decay phenomenon is expected to fail here. Indeed, according to our numerical simulations, no signs of a slower-than-exponential density decay have been found. For random regular networks, we also confirmed that the density follows the scaling behavior predicted by the mean-field theory at $\lambda_c$. 

As a possible application area of the results found in this work, we mention the paradigm of criticality in brain activity. This is manifested through power-law distributions of neuronal avalanches and has induced intense debate over the past two decades~\cite{Wilting2019} since the pioneering work of Beggs and Plenz~\cite{Beggs2003}. Self-organized criticality~\cite{Plenz2021}, brain plasticity~\cite{Levina2007}, and Griffiths phases~\cite{Moretti2013} have been proposed as possible mechanisms underlying such behavior. In the present work, we showed that reducibility of the state space, whereby transitions between distinct regions of the configuration space are forbidden, can also produce slowly decaying dynamics, a hallmark commonly associated with dynamical critical phenomena. Moreover, the brain exhibits a modular and hierarchical organization~\cite{Meunier2010a}, with different regions displaying partially autonomous dynamics, which suggests a possible analogy with the reducibility of configuration space considered here. While it is premature to claim that the mechanism investigated in this work can explain power-law behavior in the brain or other networked systems, our results reveal an alternative mechanism for the emergence of slow relaxation that does not rely on conventional criticality.

\begin{acknowledgments}
This work was supported by the National Research, Development and Innovation Office NKFIH under Grant No.~K146736.
S.C.F. acknowledges financial support from the \textit{Conselho Nacional de Desenvolvimento Científico e Tecnológico} - CNPq  (Grants No. 310984/2023-8  and 407871/2025-0), INCT-NeuroComp (CNPq Grant No. 408389/2024-9), and \textit{Fundação de Amparo à Pesquisa do Estado de São Paulo} - FAPESP (Grant No. 25/24366-1), and \textit{Fundação de Amparo à Pesquisa do Estado de Minas Gerais} - FAPEMIG (Grant No. APQ-01973-24). 
This study was financed in part by the \textit{Coordenação de Aperfeiçoamento de Pessoal de Nível Superior} (CAPES), Brazil, Finance Code 001.
\end{acknowledgments}

\appendix

\section{Growth rate of an infinitely long active cluster}
\label{app:growth}
Based on the phenomenological description of slow decay,
we investigate the growth rate of the thickness of an infinitely long active cluster that is growing out from a line of active sites with $\mu=0$. 
Initially, at places where there is an active cluster in the boundary zone of the line, a new layer starts to develop parallel to the line, with a finite growth velocity. The time needed for the formation of a complete layer is proportional to the typical distance $d$ between  clusters in the boundary zone of the growing active line, $\tau\sim d\sim 1/n_0$. The width of this first layer is given by the typical linear size $\xi$ of active clusters since the boundary of the core cluster is approximately straight, except in regions where neighboring clusters come into contact with it, forming protrusions whose typical extent perpendicular to the boundary is of order $\xi$. Then, activity spreads laterally in both directions, parallel to the original boundary, progressively filling the gaps between neighboring protrusions forming a layer of width $\xi$.
Assuming a layer-by-layer growth of the core cluster, the time at which the $n$th layer starts to form will be denoted by $t_n$. The time required for the formation of the $n$th layer is $\tau_n\sim d(t_n)\sim 1/n(t_n)$.
Clearly, we have $t_n=t_{n-1}+\tau_{n-1}$. 
For a power-law decay of the number density as in Eq. (\ref{nt}), we can write a recursion
\be 
t_n=t_{n-1}+Ct_{n-1}^{\delta},
\label{recursion}
\ee
with a constant $C$.
The width of the $n$th layer is $\xi(t_n)\sim \frac{1}{c}\ln(at_n)$.
The sequence of times $t_n$ increases with the layer index $n$ qualitatively differently depending on $\delta$. Two domains can be distinguished.

If  $\delta>1$, the second term on r.h.s. of Eq. (\ref{recursion}) is dominant over the first one, and we obtain $t_n\sim C^{1/(1-\delta)}e^{\delta^n}$ for large $n$. The width of the core cluster with $n$ layers is 
 $l_n\sim \sum_n\xi(t_n)\sim \frac{1}{\delta-1}\delta^n$. 
 Expressing $\delta^n$ by the time using the asymptotic expression for $t_n$, we obtain for the time dependence of the width 
 \be 
 l(t)\sim \ln t    \qquad (\delta>1).
 \ee

 For $\delta=1$, we have $t_n=(1+C)^nt_0$ and $l_n\simeq \frac{1}{c}\ln(1+C)\frac{n^2}{2}$, leading to
 
\be 
 l(t)\sim (\ln t)^2    \qquad (\delta=1).
 \ee
 
For $\delta<1$, to obtain the variation of $t_n$ with $n$, we may rewrite the recursion as $t_n-t_{n-1}\equiv \Delta t_{n-1}=Ct_n^{\delta}$, and approximate it by the differential equation $\frac{dt}{dn}=Ct^{\delta}$. This has the solution
$t^{1-\delta}/(1-\delta)=Cn+const$, by which we have the asymptotic behavior of $t_n$ as $t_n\sim (1-\delta)^{1/(1-\delta)}n^{1/(1-\delta)}$.
This gives for the width of the core cluster
$l_n\sim \frac{1}{1-\delta}n\ln n$, and yields for the asymptotic time dependence
\be
l(t)\sim t^{1-\delta}\ln t  \qquad (\delta<1).
\ee

\section{Triangular and higher dimensional hyper-cubic lattices}\label{app:higher}

\begin{figure*}[tbh]
    \centering
    \includegraphics[width=1\linewidth]{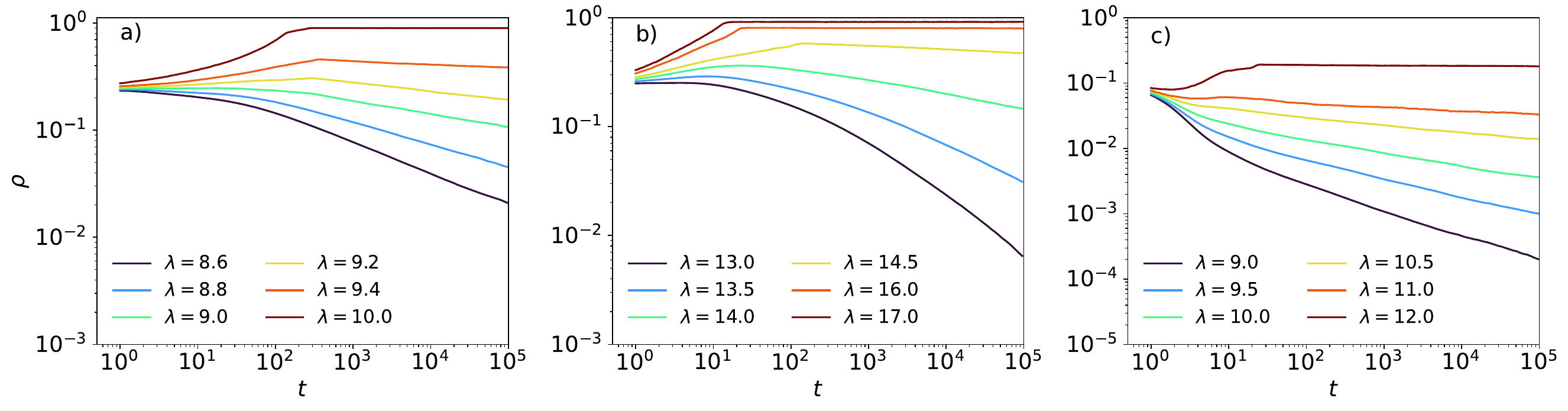}
    \caption{Time dependence of the density obtained numerically on different lattices: (a) triangular lattice of size $L=200$ and parameters $\Theta = 3/6$ and  $\rho(0)=0.2$; (b) cubic lattice of size $L=35$ and parameters $\Theta = 3/6$ and $\rho(0)=0.2$;  and (c)  four-dimensional lattice with linear size $L=20$ and parameters $\Theta = 3/8$, $\rho(0)=0.1$.  
    \label{fig:tri}
    }
\end{figure*}

We investigated the possibility of slow density decay in the threshold model on a 2d triangular
and on higher dimensional hyper-cubic lattices with periodic boundary conditions. 
The results as well as the simulation parameters are displayed in Fig.\ref{fig:tri}.
In all three cases presented here, we can observe an algebraic decay of the density with non-universal exponents.
We note that although the phenomenon of slower-than-exponential decay is quite general in the threshold model on regular lattices irrespective of the lattice type and the spatial dimension, the functional form $\rho(t)$ may be different from a power law, depending on $\Theta$, $d$ and the lattice type.   

\section{Discrete-time model}
\label{sec:AIII}

In this section, we consider a synchronous-update variant of the threshold model, introduced in~\cite{HMNcikk} to describe brain dynamics. 
The initial conditions are $\sigma_i=1$ for a fraction $\rho_0$, and a full-lattice update is carried out with time steps $\Delta t =1$, according to the following update rules. All inactive nodes that have at 
least one active contact are visited and if the condition $\sum_j A_{ij} \sigma_i \ge K \sum_j A_{ij}$
is satisfied, then the site is activated with a probability $\tilde\lambda$.
Active nodes are inactivated with a probability $\tilde\nu$.  
This process differs not only in the type of update (synchronous) from the continuous-time model. 
There is a further difference: here, vacant sites are picked and activated with the same probability 
irrespective of the number of active sites (if the threshold condition is fulfilled). 
In the continuous-time model the rate of filling of a vacant site does depend on the number of active neighbors.

We performed numerical simulations on 2d lattices with linear size $L=400$ and periodic
boundary conditions up to $10^6$ full lattice updates. Averaging was done over $100$ runs with different initial conditions.
For $K\le 1/4$, one active neighbor can activate, and we found a 2d directed percolation type of continuous
phase transition, characterized by the density decay $\rho\propto t^{-0.45}$ at 
$\tilde\lambda_c \simeq 0.317$, for $\tilde\nu=0.9$. 

\begin{figure}[ht]
\includegraphics[width=80mm]{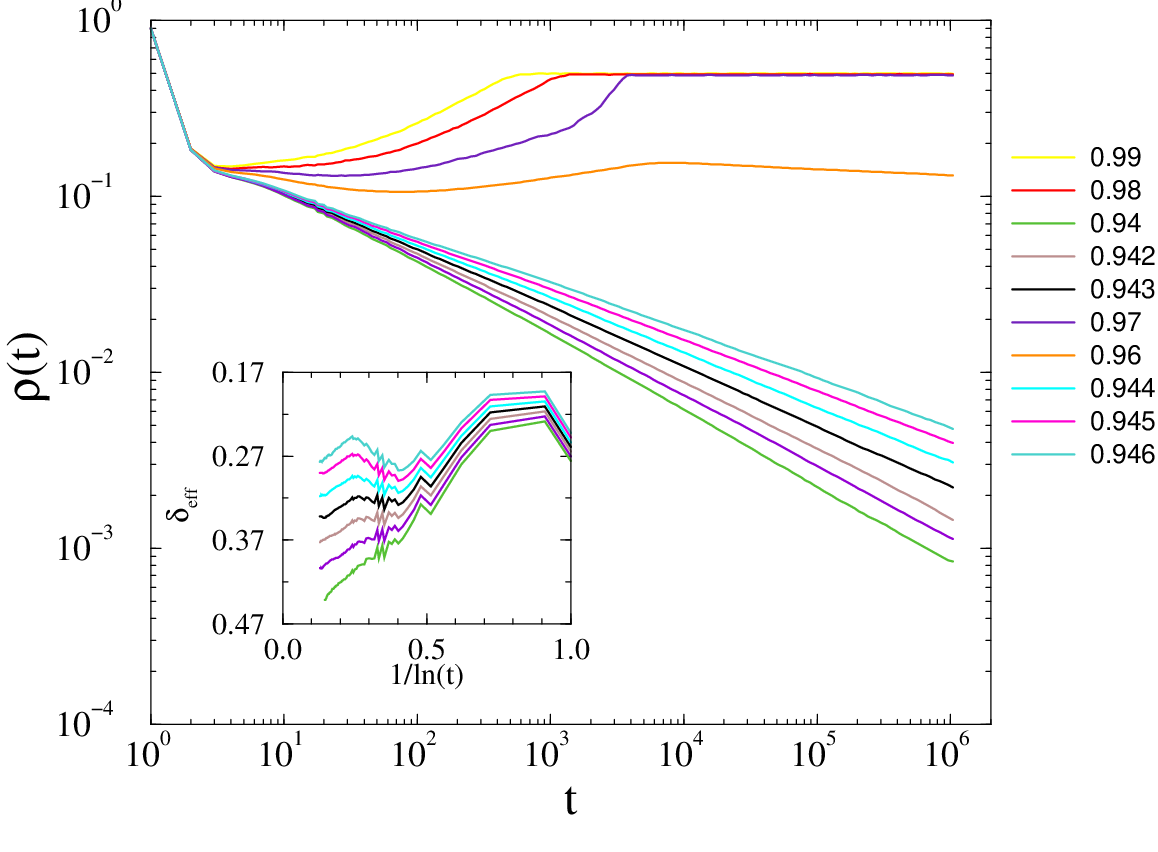}
\caption{\label{k=0.25-dec} 
Time-dependence of the density in the discrete-time threshold model on a square lattice in the case when at least two active neighbors are needed for an activation event. The initial density was $\rho(t=0)=0.9$, the deactivation probability was $\tilde\nu=0.9$.
Inset: effective exponents of the same results, suggesting non-universal power-law decay. 
}
\end{figure}

In the case of $1/4<K\le 1/2$, where at least two active neighbors are needed for an activation event, non-universal algebraic decay of the density appears at $\tilde\nu=0.9$ in a range of $\tilde\lambda$, as can be seen in 
Fig.\ref{k=0.25-dec}. 
Increasing $\tilde\lambda$, the decaying character of curves changes abruptly at $\tilde\lambda=0.969(1)$, above which the density tends to a ($\tilde\lambda$-dependent) positive limit.
The local slopes of decay, defined as 
\begin{equation}  \label{deff}
\delta_\mathrm{eff}(t) = - \frac {\ln \rho(t) - \ln \rho(t') } {\ln(t) - \ln(t')} \ ,
\end{equation}
using $t/t'= 2$ are also shown in the inset, exhibit corrections to the scaling, in agreement with 
the logarithmic corrections discussed in Section~\ref{sec:pheno}. 
This standard analysis~\cite{odorbook} allows to estimate the asymptotic values in the $1/\ln(t)\to 0$
limit on the vertical axis and allows to follow corrections to the scaling, when the leading order correction
is $O(1/\ln(t))$.
Note that most of the extrapolated values of $\delta$ obtained here are smaller than $1/2$, thus, according to the phenomenological theory in Sec. \ref{sec:stab}, the observed decay must be metastable. This means that in an infinite system, the decay of density must cross over to an increase and a saturation to a positive stationary value.  

Note, that due to the synchronous dynamics at the first full-lattice update creation is
not possible if we start from an almost full initial state, but due to the high $\tilde\nu$
the $\rho$ falls down immediately to $\simeq 0.1$. 
So, it is very similar to the random sequential dynamics simulations with low initial 
concentration of activity.

\bibliographystyle{apsrev4-2}
\bibliography{apssamp.bib}

\end{document}